\documentclass[aps,reprint,nofootinbib,superscriptaddress]{revtex4-1}

\providecommand{\sorthelp}[1]{}
 
\usepackage{graphicx}
\usepackage{amssymb}
\usepackage{epstopdf}
\usepackage{multirow}
\usepackage[colorlinks=true,linkcolor=blue,citecolor=blue]{hyperref}%
\usepackage{graphicx}
\usepackage{bm}
\usepackage{color}
\usepackage[caption=false]{subfig}
\usepackage{mhchem}

\newcommand{\be}{\begin{eqnarray}}

\newcommand{\ee}{\end{eqnarray}}

\usepackage[normalem]{ulem}

\usepackage[export]{adjustbox}

\providecommand{\sorthelp}[1]{}

\begin{document}

\preprint{APS/123-Q ED}

\title{The Atacama Cosmology Telescope: High-resolution component-separated maps across one-third of the sky}  


\author{William~Coulton}\affiliation{Center for Computational Astrophysics, Flatiron Institute, 162 5th Avenue, New York, NY 10010 USA}
\author{Mathew~S.~Madhavacheril}\affiliation{Department of Physics and Astronomy, University of
Pennsylvania, 209 South 33rd Street, Philadelphia, PA, USA 19104}\affiliation{Perimeter Institute for Theoretical Physics, Waterloo, Ontario, N2L 2Y5, Canada}
\author{Adriaan~J.~Duivenvoorden}\affiliation{Center for Computational Astrophysics, Flatiron Institute, 162 5th Avenue, New York, NY 10010 USA}\affiliation{Joseph Henry Laboratories of Physics, Jadwin Hall,
Princeton University, Princeton, NJ, USA 08544}
\author{J.~Colin~Hill}\affiliation{Department of Physics, Columbia University, New York, NY, USA}\affiliation{Center for Computational Astrophysics, Flatiron Institute, 162 5th Avenue, New York, NY 10010 USA}
\author{Irene~Abril-Cabezas}\affiliation{DAMTP, Centre for Mathematical Sciences, University of Cambridge, Wilberforce Road, Cambridge CB3 OWA, UK}\affiliation{Kavli Institute for Cosmology Cambridge, Madingley Road, Cambridge CB3 0HA, UK}
\author{Peter~A.~R.~Ade}\affiliation{School of Physics and Astronomy, Cardiff University, The Parade, 
Cardiff, Wales, UK CF24 3AA}
\author{Simone~Aiola}\affiliation{Center for Computational Astrophysics, Flatiron Institute, 162 5th Avenue, New York, NY 10010 USA}\affiliation{Joseph Henry Laboratories of Physics, Jadwin Hall,
Princeton University, Princeton, NJ, USA 08544}
\author{Tommy~Alford}\affiliation{Department of Physics, University of Chicago, Chicago, IL 60637, USA}
\author{Mandana~Amiri}\affiliation{Department of Physics and Astronomy, University of
British Columbia, Vancouver, BC, Canada V6T 1Z4}
\author{Stefania~Amodeo}\affiliation{Universit{\'{e}} de Strasbourg, CNRS, Observatoire astronomique de Strasbourg, UMR 7550, F-67000 Strasbourg, France}
% ...


\date{\today}

\begin{abstract}

Observations of the millimeter sky contain valuable information on a number of signals, including the blackbody cosmic microwave background (CMB), Galactic emissions, and the Compton-$y$ distortion due to the thermal Sunyaev-Zel'dovich (tSZ) effect. Extracting new insight into cosmological and astrophysical questions often requires combining multi-wavelength observations to spectrally isolate one component. 
In this work, we present a new arcminute-resolution Compton-$y$ map, which traces out the line-of-sight-integrated electron pressure, as well as maps of the CMB in intensity and E-mode polarization, across a third of the sky (around 13,000 sq.~deg.). We produce these through a joint analysis of data from the Atacama Cosmology Telescope (ACT) Data Release 4 and 6 at frequencies of roughly 93, 148, and 225 GHz, together with data from the \textit{Planck} satellite at frequencies between 30 GHz and 545 GHz. We present detailed verification of an internal linear combination pipeline implemented in a needlet frame that allows us to efficiently suppress Galactic contamination and account for spatial variations in the ACT instrument noise.  
These maps provide a significant advance, in noise levels and resolution, over the existing \textit{Planck} component-separated maps and will enable a host of science goals including studies of cluster and galaxy astrophysics, inferences of the cosmic velocity field, primordial non-Gaussianity searches, and gravitational lensing reconstruction of the CMB.
\end{abstract}

\pacs{Valid PACS appear here}
\maketitle

\section{Introduction}
\label{sec:intro}
Millimeter observations of the sky provide a window into the universe across cosmic history as they comprise signals from our solar system \citep{Naess_2021}, our Galaxy \citep{Baxter_2018,planck2015-XXXIV,planck2014-XVIII,planck2016-XLIV}, galaxy clusters \citep{Sunyaev_1972,Sunyaev_1980,Staniszewski2009,planck2014-a28,Hasselfield2013,Bleem2015}, high-redshift star-forming galaxies \citep{planck2013-pip56}, the cosmic microwave background (CMB) \citep{Penzias_1965,Fixsen_1996,hinshaw2012,Dutcher_2021,Aiola_2020}, and more. This profusion of signals makes these observations well suited for learning about astrophysical and cosmological processes. However, it also comes at a cost: the information from any given process is mixed with the multitude of other signatures. Sources of noise and instrumental effects further complicate these measurements. Though there are times when it may be best to deal directly with the unprocessed data sets (for instance in the analysis of the power spectrum of the primary CMB anisotropies) for many science cases it is beneficial to isolate a component of interest from others; collectively, methods to address this task are known as component separation techniques.\footnote{See \url{https://lambda.gsfc.nasa.gov/toolbox/comp_separation.html} for a collation of CMB component separation methods.} Through these methods, we can produce sky maps of components of interest with reduced contamination from the other sky signals, thereby enabling detailed studies of the relevant physical processes.

Component separation methods can be roughly divided into two categories: blind and unblind methods. In the prototypical unblind approach, a model of the sky is developed and parameters describing the scale and/or spatial and frequency dependence of the components of the sky is fit to the data; an example of this is the \textsc{commander} method used in \textit{Planck} and BeyondPlanck \citep{eriksen2006,eriksen2008,BeyondPlanck_2022}.  On the other hand, blind methods make minimal assumptions about the contributions to the observations, with the simplest methods only assuming that the frequency dependence of the component of interest is known, and focus on using the empirical properties of the data. Blind and semi-blind approaches include \textsc{fastica}, \textsc{sevem}, \textsc{smica}, \textsc{gnilc}, and \textsc{milca} \citep{maino2002a,fernandez2012,cardoso2008,Remazeilles_2011,Hurier_2013}. These approaches each have their merits; blind approaches are typically highly flexible, simple, and fast, whilst unblind approaches can provide complete models of the sky and easily include complex priors \citep[see e.g.,][for a comparison of these approaches]{Delabrouille_2007,leach2008}. 

In this work we use a blind method, known as the internal linear combination (ILC) method. Since the first application of the ILC method to the COBE data by Ref.~\citep{Bennett1992}, this method has been extensively used in the analysis of CMB data, including data from \textit{WMAP}, \textit{Planck}, the Atacama Cosmology Telescope (ACT), and South Pole Telescope (SPT) experiments \citep{bennett2003b,delabrouille2009,planck2014-a28,Madhavacheril_2020,Aghanim_2019,Bleem_2022}. The main benefits of the ILC approach are its simplicity, minimal assumptions, and flexibility. The ILC method can be applied to data in many different domains, e.g., real space, harmonic space, or a wavelet frame --- as in this work. Wavelet frames provide joint localization in real and harmonic space. Wavelets are thus well suited to analyzing CMB data where extragalactic signals are best described in the harmonic basis, and Galactic and some instrumental effects are better described in pixel space. Wavelets were first combined with ILC methods in \citep{delabrouille2009} and have since been further developed and applied to \textit{Planck} data \citep{Remazeilles_2011,Remazeilles_2013,Hurier_2013,planck2016-l04}. Our implementation follows that developed in Ref.~\citep{Remazeilles_2013}, with a key modification: a new method to mitigate the ``ILC bias". The ``ILC bias" arises as the weights used to linearly combine the individual frequency maps are obtained from the data themselves. Our mitigation method works by ensuring that these empirically determined weights are never applied to the same data from which they were estimated.

We focus on studying two sky signals: (i) the thermal Sunyaev-Zel'dovich (tSZ) effect and (ii) the blackbody component in temperature and polarization. The latter blackbody component includes the lensed CMB in temperature and polarization and the kinetic Sunyaev-Zel'dovich (kSZ) effect in temperature. The Compton-$y$ signal, sourced by the tSZ effect \citep{Sunyaev_1972,Sunyaev_1980_tsz}, is an important cosmological and astrophysical probe as it traces the distribution of free electrons, from hot ionized matter, throughout the universe. Isolating this signal from the dominant foreground signals is essential for studies as diverse as constraints on massive neutrinos or on cluster feedback processes \citep{Madhavacheril_2017,Pandey_2019,Amodeo_2021}.
Component-separated, or cleaned, blackbody temperature maps are needed for a diverse range of studies including CMB lensing and primordial non-Gaussianity analyses \citep{vanEngelen_2014,Osborne_2014,Hill_2018,Coulton_2022}, where it is important to remove contaminant signals from Galactic and extragalactic sources to avoid biased inferences, and analyses of the kSZ effect \citep{Sunyaev_1980}, where other extragalactic signals can bias measurements of cosmic gas distributions and act as large sources of noise \citep{Hill_2016,Schaan_2021}.

In this work we apply this pipeline to new data from the upcoming ACT Data Release 6 (DR6) and previous data from ACT Data Release 4 (DR4) and the \textit{Planck} satellite \citep{Fowler_2007,planck2016-l01}. The \textit{Planck} satellite's precise measurements of the large scale millimeter sky ($\gtrsim 5'$) naturally complements ACT's high resolution measurements ($\sim 1.5'$).
The main results of this work are $1.6'$ resolution maps of the tSZ, CMB temperature, and CMB E-mode polarization anisotropies, with mean white noise levels of $\sim 15\, \mu$K-arcmin in temperature. These products build on existing component-separated maps from \textit{Planck} \citep{planck2013-p06,planck2014-a11,planck2014-a12,planck2016-l04} by utilizing the high-resolution ACT data to provide improved small-scale information. This is achieved at the cost of limiting the maps to the $\sim 1/3$ of the sky observed by ACT.  However, our new maps cover $\sim 5\times$ larger sky fractions than existing high-resolution component-separated maps, such as those from Ref.~\citep{Madhavacheril_2020} and Ref.~\citep{Bleem_2022}. Further, our use of the wavelet frame is complementary to the harmonic and Fourier space method \citep[see e.g., ][for a discussion of some tradeoffs of these frames]{Atkins_2022} used in  Ref.~\citep{Madhavacheril_2020} and Ref.~\citep{Bleem_2022} and allows for a better removal of Galactic foregrounds. 

The products of this work --- including maps with $0.5'$ pixels, simulations, and auxiliary data --- will be made available on \textsc{LAMBDA} and NERSC.\footnote{The $y$-map products will be made available at publication of this paper; the blackbody map products will be made available alongside the release of the single-frequency DR6 maps}  This paper is part of a suite of ACT DR6 papers, which will include a dedicated paper describing the single-frequency maps.

This paper is structured as follows: in Section \ref{sec:DataSets} we briefly describe the data used in this paper and in Section \ref{sec:pipeline} we provide the details of our component separation pipeline. We present the component-separated maps in Section \ref{sec:maps} and discuss a few of their key properties in Section \ref{sec:map_properties}. We then conclude in Section \ref{sec:discussion}. In Appendix \ref{app:simulations} we describe the simulations used to validate our tools and simulated products that accompany this work and in Appendix \ref{app:HarmonicILC} we describe the harmonic ILC method used as a baseline, comparison method. In Appendix \ref{app:needlet_m_modes} we provide a detailed description of the ILC bias reduction method. Finally in Appendix \ref{app:systematics} we describe how we include instrumental systematic effects into our analysis pipeline.

\section{Data Sets}\label{sec:DataSets}
The single frequency maps used in this work are from the ACT DR4 and upcoming DR6 data sets and the \textit{Planck} NPIPE analysis. Tables \ref{tab:dataSets} and \ref{tab:dataSetsPlanck}
provide summaries of the key properties of the ACT and \textit{Planck} data sets, respectively.  

The Atacama Cosmology Telescope was a 6\,m off-axis Gregorian telescope \citep{Fowler_2007}  located at an elevation of 5190\,m in the Atacama Desert of Chile, used to measure the CMB from 2007 to 2022 \citep[e.g.,][]{dunner/etal:2013,dunkley/etal:2011,das/etal:2011,sievers/etal:2013,gralla/etal:2020,louis/etal:2017,naess/etal:2014,Aiola_2020,naess/etal:2020}.  The DR4 and DR6 data comprise multifrequency observations across $\sim 1/3$ of the sky, measured by polarization-sensitive arrays of feedhorn-coupled transition-edge sensor (TES) bolometers \citep{grace/etal:2014, datta/etal:2014, ho/etal:2016, Henderson_2016, ho/etal:2017,choi/etal:2018, li/etal:2018}.  The arrays were cooled to 100~mK in a receiver providing separate optics chains (lenses and filters) for each array \citep{Thornton_2016}.  We label the detectors according to the approximate centers of their frequency responses in GHz as follows: f090, f150, and f220. The ACT maps are produced in the \textit{Plate-Carr\'{e}e} (hereafter abbreviated CAR) projection scheme. This scheme is used for both the input maps and the needlet maps. The CAR maps have a rectangular pixelization with the $x$ and $y$ axes aligned with right ascension and declination, respectively.

\begin{figure*}
    \centering
  \includegraphics[width=.99\textwidth]{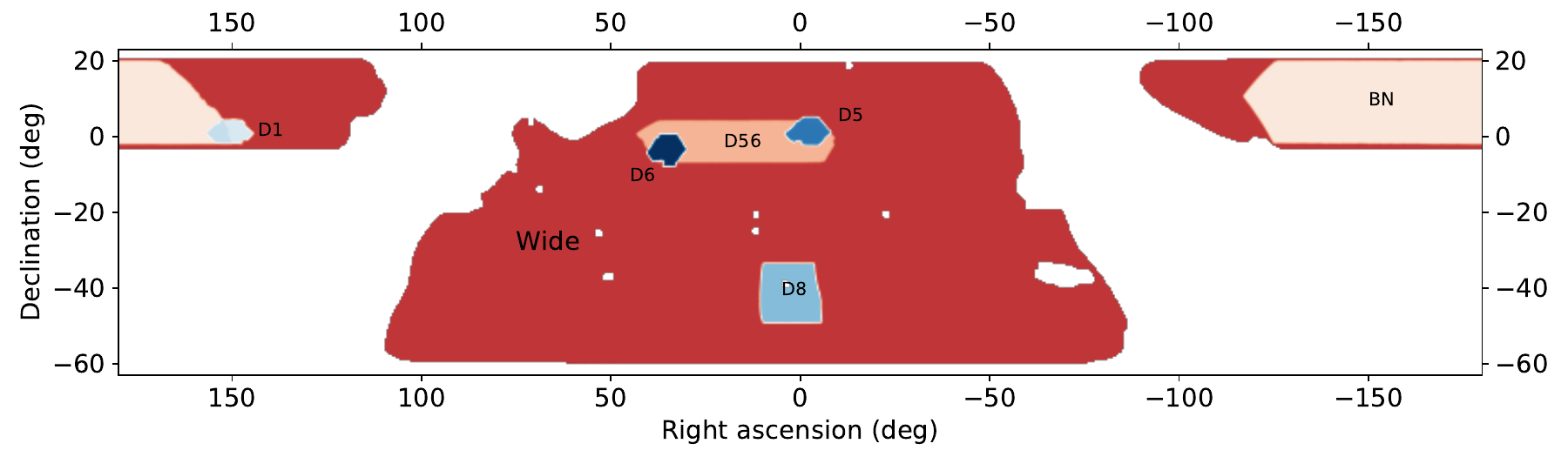}%
 \caption{ Footprints of the different data sets used in this work. ACT DR4 primarily focused on observing the  deep patches, denoted by ``D" and ``BN''. Since 2016, ACT used upgraded detectors to observe significantly wider areas, denoted by ``wide", to approximately similar depth. We use the subset of \textit{Planck} data that lies within this ``wide" region. Note that the full ACT data set extends into the Galactic plane and in this work we adopt a smaller footprint to avoid contamination from bright Galactic emissions. The excised regions within the main footprint correspond to extended sources that are also masked or inpainted. 
 \label{fig:mask_footprints} }
\end{figure*}

The ACT DR4 data cover night-time ACT observations\footnote{Night-time data are those  data taken between 23 and 11 UTC} over four observing seasons from 2013 to 2016 \citep{Aiola_2020,Choi_2020}. The DR4 data set comprises a set of deep observations in the regions labeled by ``D5", ``D6", ``D56", ``D1", ``D8", and ``BN" in Fig.~\ref{fig:mask_footprints}, as well as shallower observations of the ``wide" region. In this work, we only use the deep observations from DR4 as the noise levels in its wide field maps are too large to provide noticeable improvements in our analysis. 
The DR4 data were collected by the three arrays of the ACTPol camera \cite{Thornton_2016}. 
 The first two arrays, called PA1 and PA2 (where PA is an abbreviation for Polarimeter Array), were sensitive to the  f$150$ band (124--172$\,$GHz)\footnote{This range encompasses roughly 95\% of the area under the filter response curve;  see Fig.~\ref{fig:passbands_wErrors}.}  whilst the third array was dichroic, observing in both the f$090$ (77--112$\,$GHz) and f$150$ bands.

The DR6 data sets include observations from 2017 to 2022 at three frequency bands: f090, f150,  and f220 (182--277$\,$GHz). The observational program targeted the ``wide" field.  For this work, we use only the night-time portion of the data taken in the first five observing seasons, up to 2021.  The Advanced ACT camera, used for these observations, was equipped with three dichroic arrays: PA4 at f$150$ and f$220$, and PA5 and PA6, each sensitive to both f$090$ and f$150$ \citep{Henderson_2016}.  Each frequency band of each array was mapped separately, but observations from different seasons were combined. Each of these data sets (i.e. the separate data from each frequency of each array) was further divided into eight sub-portions, hereafter referred to as ``splits", with independent instrumental and atmospheric noise. 
This analysis uses the first science-grade version of the ACT DR6 maps, labeled dr6.01. Since these single-frequency maps were generated, some refinements have been made to the ACT mapmaking that improve the large-scale transfer function and polarization noise levels. A second version of the input maps are expected to be used for further science analyses and for the DR6 public data release, and we will update the derived products as those maps are produced and released. More details of these maps will be provided in an upcoming paper. 

In addition to the frequency maps, we used the DR4 and DR6 beams, point source catalogs, and passbands. The DR4 products are described in Ref.~\citep{Datta_2019,Aiola_2020,Gralla_2020,Lungu_2022}.\footnote{These products are publicly available at \url{https://lambda.gsfc.nasa.gov/product/act/actpol_prod_table.html}}  The DR6 products are produced by similar methods, which will be detailed in an upcoming publication. In Appendix \ref{app:systematics} we plot the DR6 passbands used in this work (Fig.~4 of Ref.~\citep{Madhavacheril_2020} shows the DR4 passbands); these are key inputs for the component separation pipeline. The ACT point source catalogs are created for each frequency by applying a matched filter to a map obtained from combining the individual data splits \citep{Stetson_1987}.

For the \textit{Planck} data, we use the NPIPE maps described in \citep{planck2020-LVII}. These single-frequency maps cover the full sky, though we only use the data within the ACT ``wide" footprint shown in Fig.~\ref{fig:mask_footprints}. \textit{Planck} has  nine frequencies ranging from $30$\,GHz to $857$\,GHz, with resolutions ranging from 32 arcmin to 4.2 arcmin. The data are provided in two splits that are independently processed and largely statistically uncorrelated.  Unlike the other frequencies, the \textit{Planck} 857 GHz channel is not calibrated on the orbital dipole and instead uses a planetary absolute calibration. The challenges and uncertainties associated with this can impact the component separation; to avoid this, we do not use the 857 GHz data. In addition to the frequency maps, we use measurements of the \textit{Planck} passbands \citep{zonca2009,planck2013-p02,planck2013-p03d,Hivon_2017} and beams \citep{planck2014-a05,planck2014-a08}.  
We compare our results to component-separated maps produced by the \textit{Planck} collaboration, specifically the MILCA Compton-$y$ map, and the \textit{Planck} NILC Compton-$y$ and CMB maps \citep{planck2014-a28,Hurier_2013,planck2016-l04}.

\begin{table}[]
    \centering
    \begin{tabular}{c|c | c| c | c |c }
    Patch & Area & Frequency & Typical Depth & FWHM &  Number of  \\
    Name  & (deg$^2$)&  Band & ($\mu$K arcmin) & (arcmin) & Data sets \\ \hline
     Wide  &  $\sim 12200$  & f090 & 20&  2.1 & 2   \\ 
     Wide  &  $\sim 12200$  & f150 & 20&  1.4 & 3   \\ 
     Wide  &  $\sim 12200$  & f220 & 65&  1.0 & 1  \\ 
     D1  &  $\sim60$  & f150 & 15 &  1.4 & 1   \\ 
     D5  &  $\sim60$  & f150 & 12 &  1.4 & 1   \\ 
     D6  &  $\sim60$  & f150 & 10 &  1.4 & 1   \\ 
     D56  &  $\sim560$  & f150 & 20&  1.4 & 5   \\ 
     D56  &  $\sim560$  & f090 & 17&  2.0 & 1   \\ 
     BN  &  $\sim1800$  & f150 & 35&  1.4 & 3   \\ 
     BN  &  $\sim1800$  & f090 & 33&  2.0 & 1   \\ 
     D8  &  $\sim 200$  & f150 & 25&  1.4 & 3   \\ 
     D8  &  $\sim 200$  & f090 & 20&  2.0 & 1   \\ 
    \end{tabular}
    \caption{A summary of key properties of the ACT data set used in this analysis. The depth here refers to the white noise level and does not include atmospheric contributions, which are dominant at large scales. Note that the full-width half-maxima (FWHM) of the beams are from a Gaussian fit; however we use the full beam profiles in our analysis. In DR4 different observation seasons and different detector arrays were mapped separately. In DR6 different arrays and different frequencies were mapped separately, but all the observation seasons were combined. This leads to multiple data sets with similar footprints and noise levels, but different passbands and beams. The number of data sets for each patch is listed in the last column. 
    \label{tab:dataSets}}
\end{table}

\begin{table}[]
    \centering
    \begin{tabular}{c | c| c | c  }
    Reference & Frequency & Typical Depth & FWHM   \\
    Name  & GHz & ($\mu$K arcmin) & (arcmin) \\ \hline
     P01 & 28.4 & 150 & 32 \\
     P02 & 44.1 & 162 & 28 \\
     P03 & 70.4 & 210 & 13 \\
     P04 & 100 & 77.4 & 9.7 \\
     P05 & 143 & 33 & 7.2 \\
     P06 & 217 & 46.8 & 4.9 \\
     P07 & 353 & 153 & 4.9 \\
     P08 & 545 & 1049 & 4.7 \\
    \end{tabular}
    \caption{A summary of key properties of the \textit{Planck} data set used in this analysis, adapted from Table 4 of Ref.~\citep{planck2016-l01}. The \textit{Planck} maps cover the full sky and two splits are provided for each frequency. Note that the typical depths are approximate as we use the NPIPE \textit{Planck} release, which has slightly lower white noise levels. 
    \label{tab:dataSetsPlanck}}
\end{table}

\section{Component Separation Pipeline}\label{sec:pipeline}
The component separation pipeline used in this work is composed of five main steps. We first outline these steps before describing the details of each stage in the remainder of this section.
\begin{enumerate}
\item Pre-processing: Before we can analyze the input frequency maps, we first perform a set of pre-processing steps. The aims of this step are: 1) to convolve the input maps to a common beam, 2) to filter the maps to remove contaminants such as scan-synchronous pickup,  and 3) to remove bright sources, which can pose challenges to component separation and leave artifacts in the output maps. These steps are very similar to those performed in the lensing analysis, described in Ref.~\citep{Qu_2022}. 
\item Needlet Decomposition: The next step is to transform the input maps into the wavelet frame. In this work we use the generalized ``needlet'' kernels \citep{Narcowich_2006,Guilloux_2007}; hence we refer to this frame as the needlet frame. The decomposition is achieved by convolving the input maps with the needlet kernels. This is implemented as a series of spherical harmonic transforms and filtering operations. 
\item Component Separation: At each needlet scale we apply our component separation method --- the needlet frame internal linear combination (NILC) method. This combines all the measurements at each needlet scale into a map of the component of interest. Using the methods developed in Ref.~\citep{Remazeilles_2011b}, we additionally generate maps of specific components that have other components removed (e.g., CMB temperature maps that are explicitly constructed to contain no tSZ anisotropies).  
\item Inverse Needlet Decomposition: We then transform the ILC output from the needlet frame into the real-space basis. This is achieved by reconvolving the maps with the needlet kernels.
\item Correction for mode filtering: During the preprocessing of the ACT maps we apply a filtering step that removes a set of modes from the ACT maps. The aim is to remove modes contaminated by scan-synchronous pickup. Whilst this filtering step is not performed on the \textit{Planck} maps, the absence of these modes in the ACT maps means they are partially missing from the output component-separated maps. To account for this we perform a final correction step. This step replaces the missing modes with those from a component-separated map formed from only the \textit{Planck} maps. 
\end{enumerate}
\subsection{Preprocessing}
We preprocess the \textit{Planck} and ACT maps in slightly different manners that are detailed below. The methodology for this closely follows that from Ref.~\citep{Aiola_2020} and Ref.~\citep{Madhavacheril_2020}. 
\subsubsection{\textit{Planck} preprocessing}

We perform five preprocessing steps: first we project the \textit{Planck} maps from their native HEALPix to CAR projection. Whilst the component separation pipeline does not require CAR maps, this step simplifies various preprocessing steps, such as the use of common masks. Next, we subtract sources from the two NPIPE splits at each frequency.  
For each data split we find the amplitudes of all point sources present in either the ACT or the \textit{Planck} point source catalogs. As described in \citep{Qu_2022}, the ACT catalogs are made by running a matched filter on a version of the Data Release 5 ACT+\textit{Planck} maps \citep{naess/etal:2020} updated to use the new data in DR6, and objects detected at greater than $4\sigma$ are added to the catalog for each frequency band. We account for overlapping sources in this fit. We fit the amplitude to each data split to partially account for source variability (e.g., between the measured ACT flux and the brightness of the source in the \textit{Planck} maps). We then select all the point sources whose amplitudes are detected at $>2\sigma$ in the \textit{Planck} data and subtract a model of these from each input map.  The model is given by the real-space beam profile scaled by the appropriate amplitude. Second, we mask the map; the mask we use for the \textit{Planck} data is composed of three parts: a Galactic mask (the \textit{Planck} {70\%} mask that masks the center of the Galaxy and leaves $70\%$ of the sky unmasked), a footprint mask that bounds the region observed by ACT, and a mask that removes bright extended sources from the maps, which otherwise can cause artifacts in an analogous manner to the point sources. We masked $\sim 20$ extended sources and we refer the reader to Ref.~\citep{Qu_2022} for more details on the construction of the mask. We coadd the two splits in real space using an approximate inverse variance noise weighting, based on the per-pixel inverse variance maps. 

The final preprocessing step is to transform the map to harmonic space and convolve the map to a common beam with a 1.6 arcmin full-width half-maximum (FWHM). We do this by applying an $\ell$-dependent weight given by the ratio of the harmonic transform of a 1.6 arcmin FWHM Gaussian beam to the harmonic transform of the NPIPE beams. The 1.6 arcmin FWHM is chosen to match the ACT resolution and is much smaller than the \textit{Planck} FWHMs. This means that on small scales this ratio can become very large. To avoid numerical artifacts, and reduce the computational cost, we apply a small-scale cut that excludes scales where this ratio is $\gtrsim 50$. These scales are noise dominated and have no weight in the component separation pipeline. Thus, the results are insensitive to the specific value of this cut.

\subsubsection{ACT preprocessing}

The preprocessing of the ACT maps consists of five key steps: first, in a manner identical to the \textit{Planck} maps, we subtract models for all the bright sources from the input map. The fit is performed for each split and, in this case, we remove all sources detected at or above 5$\sigma$. A lower threshold can be used for \textit{Planck} as sources  with  $\mathrm{SNR}\leq 5\sigma$ in \textit{Planck} are detected at much higher significance in the ACT maps. Next we perform an inpainting step: at the location of all the point sources detected above $70 \sigma $ we mask the region with a hole of radius 6 arcmin and fill in the masked region with a constrained realization \citep{Bucher_2012}. The aim of this is to remove large residuals that arise from the imperfect source subtraction. A larger inpainting, 10 arcmin radius, is also performed for all the bright extended, non-SZ, objects (such as bright extended radio jets) that are detected at more than $10\,\sigma$. These objects are the same as those masked in \citep{Qu_2022}. The \textit{Planck} maps are not inpainted for two reasons: 1) it is computationally expensive; and 2) the scales where the residuals are non-trivial have minimal weight in the output maps as those scales are better measured by ACT. These two inpainting operations reduce the total observed area of the ACT maps by $0.3\%$. We then coadd the splits in real space, using an approximate inverse noise weighting constructed from the per-pixel inverse variance maps. We apply a small harmonic-space correction to each split to account for the small differences in the beams between the splits.

As is described in detail in Ref.~\citep{Louis_2017,Choi_2020,Mallaby_Kay_2021}, data from ACT typically are filtered at the map-level with a Fourier-space filter. This filter is used to remove noisy modes and, most importantly, scan-synchronous pickup. The statistical properties of the scan-synchronous pickup are hard to model and so accurately understanding how they impact analyses is not feasible. To avoid biases in many analyses, these modes need to be filtered out. The scan-synchronous pickup is approximately fixed with respect to the ground. Through ACT's constant-elevation scans, the scan-synchronous pickup is then projected to horizontal stripes in the CAR maps that are well described by a small number of Fourier modes. In this work we remove the contaminated modes using the following Fourier-space filter:
\begin{align}\label{eq:kspaceFilter}
f(\mathbf{k}) &=1-\nonumber \\ &  \exp\left[{-\frac{1}{2} \left(\frac{k_x}{k^\mathrm{filter}_x}\right)^2 }\right]\left(1+\exp\left[\frac{|k_y|-k_y^\mathrm{central})}{k_y^\mathrm{width}} \right]\right)^{-1} ,
\end{align}
where $k^\mathrm{filter}_x$ controls the width of filtering in the $k_x$ directions and $k_y^\mathrm{central}$  and $k_y^\mathrm{width}$ regulate the maximum scale impacted by the filter. The values we use here are $k^\mathrm{filter}_x=5$, $k_y^\mathrm{central}=1250$  and $k_y^\mathrm{width}=50$. 

The filter applied here is different from that of previous analyses of ACT data \citep[e.g.,][]{Louis_2017,Choi_2020,Qu_2022}.  Our filter removes less small-scale power and tends to avoid large artifacts around point sources and clusters. Our filter is not designed to remove all contaminated modes; instead the primary aim of this initial filtering is to mitigate the impact of the large noise in these modes. These very noisy modes will result in a suboptimal needlet ILC map, as the isotropic needlets used here cannot deal with the Fourier space anisotropy. There are two key reasons why we do not attempt to completely remove the contaminated modes: first, the scan-synchronous pickup modes are naturally suppressed as we are combining our data with external data sets; these modes are not present in the \textit{Planck} data set and so are treated like effective noise and therefore down-weighted. Second, these contaminated modes do not impact cross-correlation based analyses, which are anticipated to be one of the main science applications of these maps. For cross-correlations it is better to have maps with simple properties (e.g., without missing modes or strongly anisotropic noise that can arise from the filtering). If contamination from scan-synchronous pickup is a potential concern, then the final NILC maps should be filtered more aggressively.  As discussed in Section \ref{sec:filteringCorrection}, we apply a correction for this filtering and replace the removed modes with those from \textit{Planck}.

The final step is to then transform to harmonic space and apply a series of harmonic-space weights. We first convolve the maps to the common beam with 1.6 arcmin FWHM, where we use the harmonic transforms of the frequency-dependent ACT beams at the central frequency. Note that we will account for frequency-dependent beam corrections at a later stage in the analysis. In Ref.~\citep{Choi_2020} the low-$\ell$ temperature multipoles were removed from the ACT maps due to an observed lack of power. It has been found, for example through a comparison with \textit{Planck} observations, that there is a scale-dependent loss of power in the ACT maps. The full origin of this feature is not known but Ref.~\citep{Naess_2022} explored how modeling errors, such as subpixel effects, can lead to similar biases. A second effect, which is thought to contribute to the loss of power, is from small inconsistencies in the individual detector gains. We quantify this effect using ``transfer functions'', obtained from fitting smooth functions to the ratio of the ACT auto- to ACT $\times$ \textit{Planck} power spectra. We deconvolve these from the ACT maps as the final preprocessing step. Empirically, the transfer functions only appear on large scales, where atmospheric noise becomes a major contribution to the map's auto-power spectrum. To ensure that we only use the maps where the fits to the transfer function are accurate, we only include ACT data on scales where the transfer function is measured to be $>95\%$. We retain scales with $\ell\gtrsim 400, 600,$ and $1000$ for f090, f150, and f220 respectively in temperature. Note that at higher frequencies, where the atmospheric noise is larger, the transfer function is important down to smaller scales. A more detailed discussion of these transfer functions will be provided in the upcoming power spectrum analysis of these maps.  As with \textit{Planck} we exclude small scales where the ratio of the common 1.6 arcmin FWHM beam to the map beam is $\gtrsim 50$. This primarily affects the $90$\,GHz data.
\subsection{Needlet Decomposition}
Wavelets are a useful frame to represent the data as they allow joint localization in real and harmonic space. These properties mean wavelets are well suited for component separation where the sky components vary significantly both spatially and with scale. There are a wide range of different types of wavelets and in this work we use the set of axisymmetric wavelets known as needlets. Needlets were developed in \citep{Narcowich_2006,Starck_2006,Guilloux_2007,Marinucci_2008} and we refer the interested reader to those papers for more details. An alternative set of scale-discrete wavelets was used to model the ACT DR6 noise properties in \citep{Atkins_2022}.

\begin{figure}
    \centering
  \includegraphics[width=.45\textwidth]{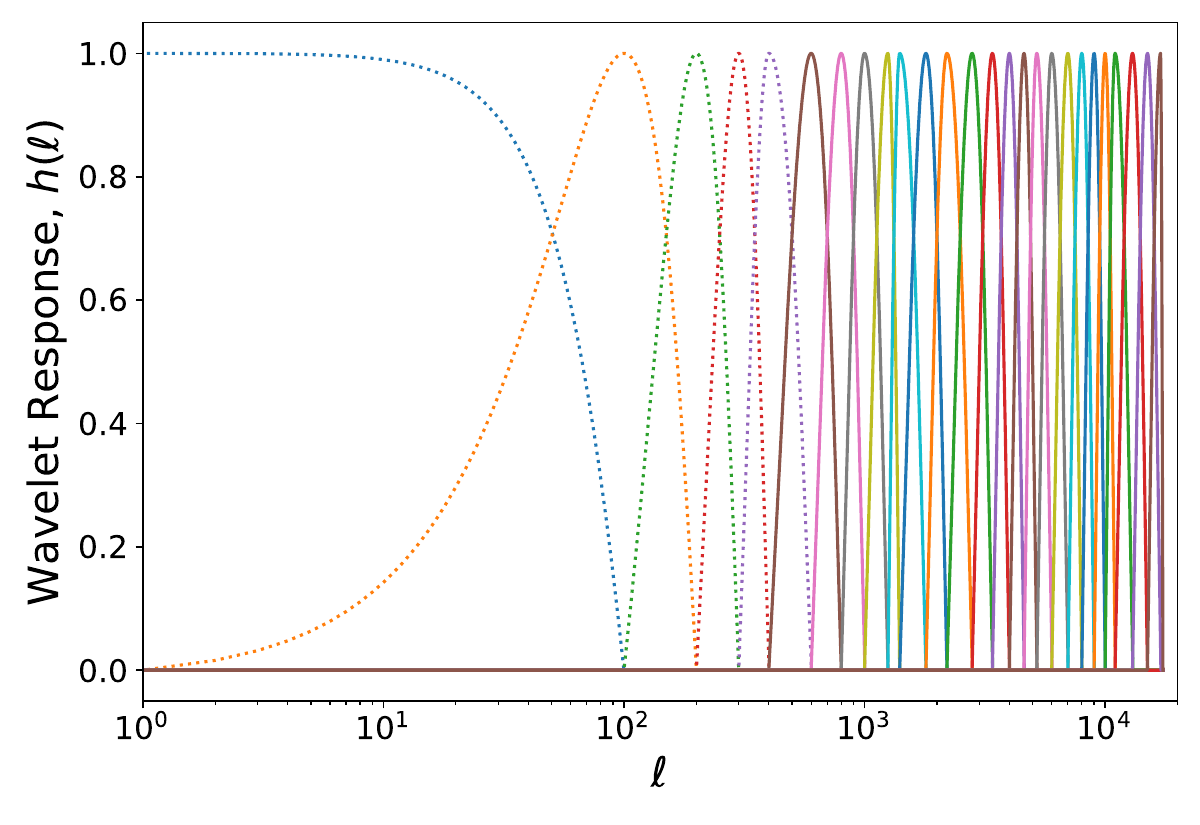}%
 \caption{Needlets allow signals to be localized in both real and harmonic spaces. Here we plot the spectral bands used to define our needlets. Wide harmonic-space bands provide better spatial localization, whilst narrow harmonic-space bands enable better separation of signals with different scale, $\ell$, dependence. We balance these two aspects by tuning the width of the bands based on the expected properties of the sky signals.  The colors are to aid differentiating one kernel from another. The dotted lines indicate scales that only include \textit{Planck} data. \label{fig:needlet_scales} }
\end{figure}

Performing a needlet analysis involves convolving the input map with a set of needlet kernels. Each needlet kernel has finite support in harmonic space and can be defined by its spectral function, $h^{(i)}(\ell)$. In this work we use the following functional form:
\begin{align}\label{eq:needletKernel}
    h^{(i)}(\ell)= \begin{cases}  
    & \cos\left(\pi \frac{ \ell^{(i)}_\mathrm{peak}-\ell}{\ell^{(i)}_\mathrm{peak}-\ell^{(i-1)}_\mathrm{peak}} \right) \,\text{ if } \ell^{(i-1)}_\mathrm{peak}\le\ell<\ell^{(i)}_\mathrm{peak} \\
    & \cos\left(\pi \frac{ \ell-\ell^{(i)}_\mathrm{peak}}{\ell^{(i+1)}_\mathrm{peak}-\ell^{(i)}_\mathrm{peak}} \right)  \,\text{ if } \ell^{(i)}_\mathrm{peak}\le\ell<\ell^{(i+1)}_\mathrm{peak}\\
    & 0  \text{ otherwise. }
    \end{cases}
\end{align}
In Fig.~\ref{fig:needlet_scales} we show spectral functions defining the needlets used in this work. The number and size of the needlet scales is set by $\ell_\mathrm{peak}$, defined in Eq.~\ref{eq:needletKernel}.  We use $\ell_\mathrm{peak}=$ 0, 100, 200, 300, 400, 600, 800, 1000, 1250, 1400, 1800, 2200, 2800, 3400, 4000, 4600, 5200, 6000, 7000, 8000, 9000, 10000, 11000, 13000, 15000, and 17000. These properties were determined through combining our knowledge of the key map signals with tests on simulations. Fewer and broader, in harmonic space, needlets can be used at small scales as the key signals -- the tSZ, CIB, point sources, and noise -- vary slowly as a function of scale. At large scales the Galactic signals vary rapidly with scale and spatial location, necessitating a balance between narrow harmonic-space bands (to capture the scale variation) with narrow spatial localization (to capture spatial variation). Through the convolution operations we produce a set of maps encoding the spatial variation of the modes within the harmonic band of that needlet kernel.

In this work we implement the convolutions through spherical harmonic transforms (SHTs). Thus, transforming an input map, $d(\mathbf{n})$, into a map at needlet scale $i$, $m^{(i)}(\mathbf{p})$, can be achieved via
\begin{align}
m^{(i)}(\mathbf{p}) = \sum\limits_{\ell m} Y_{\ell m}(\mathbf{p}) h^{(i)}_\ell \sum\limits_\mathrm{n} w(\mathbf{n})Y^*_{\ell m}(\mathbf{n})d(\mathbf{n}), 
\end{align}
where $ Y_{\ell m}(\mathbf{n})$ are the spherical harmonic functions and $w(\mathbf{n})$ are the per-pixel integration weights. A key feature to note is that the pixellations of the input map, denoted by $\mathbf{n}$, and the needlet map, denoted by $\mathbf{p}$, do not need to be the same. Likewise the pixel size can be different for each of the different needlet scales. This is useful as it allows needlets focused on large scales to have coarser pixellations, which dramatically decreases the computation time and memory footprint of the analysis. Whilst the input maps are at $0.5$ arcmin resolution, we use larger pixel scales for each of the needlet maps. These are chosen to be the largest pixel size that supports the band-limited signals at that needlet scale and allows for the computation of variance maps, described below, without aliasing effects. 
\subsection{Component Separation}\label{sec:component_separation}
Once in the needlet frame we apply the internal linear combination component separation method. In this section we briefly overview the ILC method and then describe the details of our implementation including: how we estimate the empirical covariance matrix, how we mitigate ILC bias, and how we account for the frequency dependence of the beams.
\subsubsection{Internal Linear Combination Method} \label{sec:ILC_intro}
The ILC method is a highly versatile component separation method that has been applied to a wide range of data sets \citep{Tegmark_1996,Tegmark_2003,bennett2003b,delabrouille2009,planck2014-a28,Madhavacheril_2020,Aghanim_2019}.  The method works by modeling the data observed at a set of frequencies, $d_\nu(\mathbf{x})$, as 
\begin{align}\label{eq:ilcSetup}
d_\nu(\mathbf{x}) = r^0_\nu s(\mathbf{x}) + n_\nu(\mathbf{x}), 
\end{align}
where $r^0_\nu$ is the response function that describes how the signal, $s(\mathbf{x})$, contributes to the sky at frequency $\nu$; and $n_\nu(\mathbf{x})$ is the noise term, which contains both instrumental noise and all other sky signals. The label $\mathbf{x}$ is a general label indicating the indexing in the chosen base --- thus $\mathbf{x}$ could represent a spatial index for a pixel-space ILC, or the $\ell,m$ index for a harmonic ILC or, as in our case, the needlet frame. The assumptions of the ILC are that the response function is known perfectly, which is generally the case for the signals discussed here, and that the signal is uncorrelated with the noise terms.\footnote{For signals such as the tSZ, the latter assumption is only approximately true as it is correlated with other sky signals. We can account for this by explicitly modeling these terms.} The ILC solution is a linear combination of the observations to obtain a reconstruction of the signal, i.e.,
\begin{align}\label{eq:basic_ilc}
\hat{s}(\mathbf{x})  = \sum_i w_{\nu_i} (\mathbf{x}) d_{\nu_i}(\mathbf{x}),
\end{align}
where the weights, $ w_{\nu_i} (\mathbf{x})$, are obtained by minimizing the reconstructed signal's variance subject to the condition that the ILC has unit response to the signal of interest, i.e.,
\begin{align}\label{eq:weightConstraint}
\sum_i w_{\nu_i}  r^0_{\nu_i}=1.
\end{align}
The solution for the weights is given by
\begin{align}\label{eq:basic_ilc_weights}
\mathbf{w} = \frac{  \mathcal{C}^{-1}\mathbf{r}^0}{\mathbf{r}^0  \mathcal{C}^{-1}\mathbf{r}^0},
\end{align}
where $ \mathcal{C}$ is the covariance matrix between observations at the different frequencies and we introduce the vector notation $\mathbf{r}^0$ for the vector of responses across frequencies. A detailed description of how the covariance matrix is computed is provided in Section \ref{sec:ILCBias}.

Whilst this solution minimizes the `noise' on the reconstructed signal, it imposes no constraints on what can contribute to this `noise'. In general this `noise' will be composed of instrumental noise and residual contaminant signals. These residual contaminants can potentially bias inferences made with the reconstructed signal and thus it is often of interest to impose additional constraints that force specific contaminants to zero. This technique was developed in Ref.~\citep{Chen_2009} and Ref.~\citep{Remazeilles_2011b} and is known as the constrained ILC. This method decomposes the noise term in Eq.~\ref{eq:ilcSetup} into a set of $N$ contaminants, $c^i(\mathbf{x})$, with known responses, $r^{i}_{\nu}$, and a residual noise term, $\tilde{n}$.  Thus the observations are given as
 \begin{align}
\mathbf{d}(\mathbf{x}) = \mathbf{r}^{0} s(\mathbf{x}) + \sum_{i=1,N} \mathbf{r}^i c^i(\mathbf{x})+ \mathbf{\tilde{n}}(\mathbf{x}) ,
\end{align}
where the different frequencies are represented by the vector notation.
A linear combination of the data vector is constructed as before; however, in addition to the constraints of unit response and minimum variance, $N$ additional constraints are imposed such that there is zero response to the contaminants, i.e., $\sum_i w_{\nu_i} r^{\alpha}_{\nu_i} = 0$ for $\alpha \in \{1...N\}$.  A compact form of the weights in this general case is
\begin{align}
\mathbf{w} =  \mathcal{C}^{-1} \frac{1}{\det{\mathcal{Q}}} \sum\limits_\alpha (-1)^{\alpha} \det \mathcal{Q}^s_\alpha \mathbf{r}^{\alpha}, 
\end{align}
where 
\begin{align}
\mathcal{Q}_{ij} = \mathbf{r}^{i}_{\nu_a} \mathcal{C}^{-1} \mathbf{r}^{j}_{\nu_b}
\end{align}
contains the mixing of the different components --- note that the zeroth component refers to the signal of interest --- and $\mathcal{Q}^s_\alpha $ is the matrix obtained after removing the $\alpha$th row and the zeroth column. This expression, from Ref.~\citep{Kusiak_2023}, is a refactoring of the result given in Ref.~\citep{Remazeilles_2020}.

\subsubsection{ILC bias}\label{sec:ILCBias}

\begin{figure}
    \centering
  \includegraphics[width=0.48\textwidth]{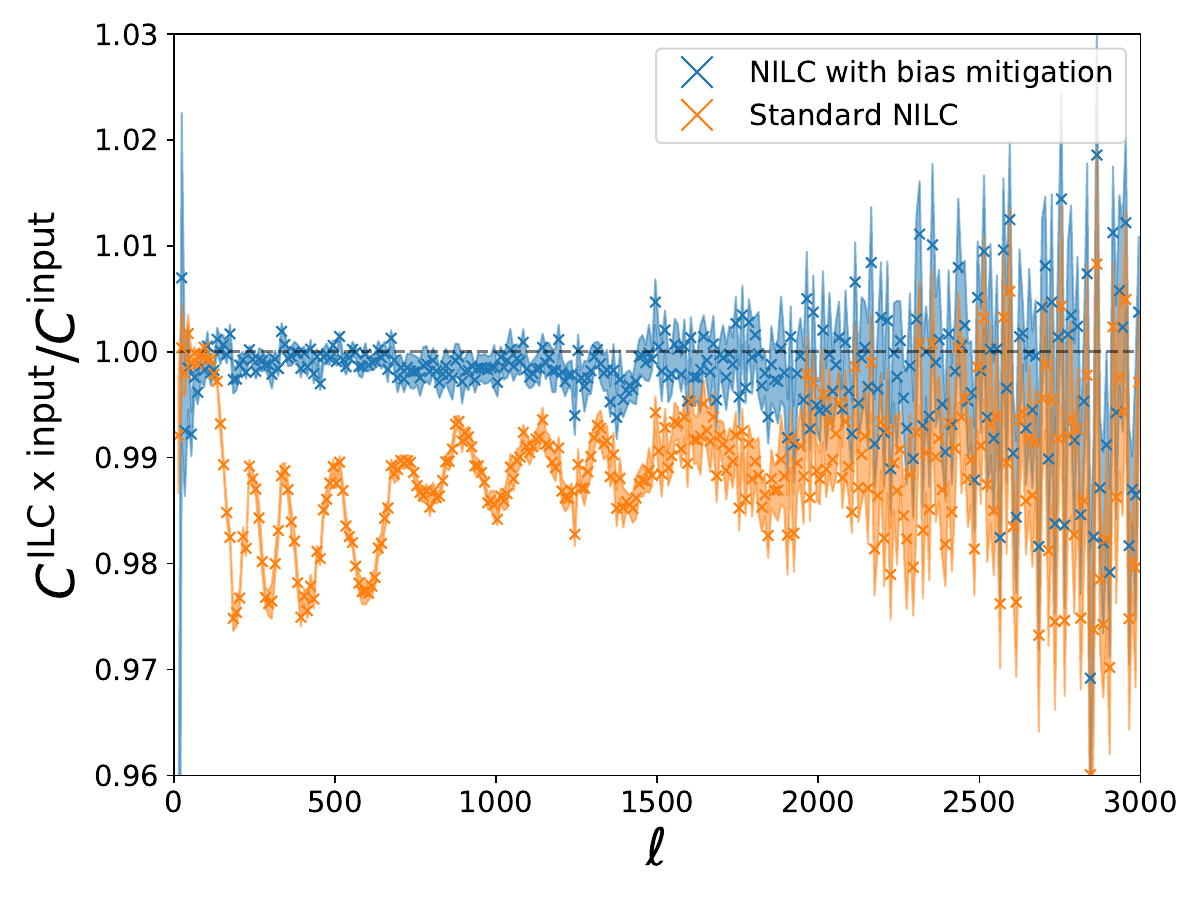}%
 \caption{The ILC method requires a frequency-frequency covariance matrix, which is typically not known a priori and therefore must be estimated from the data. This double use of the data, in both the weights and maps in Eq.~\ref{eq:basic_ilc}, leads to biases in the ILC map. In simulations, this bias can be seen by computing the cross-correlation of the ILC map with the input component map. The orange line shows the size of this ILC bias in our needlet pipeline when analyzing simulations of a subset of ACT and \textit{Planck} maps. The shaded contours denote the error on the mean. In this work we introduce a novel mitigation strategy, described in Section \ref{sec:ILCBias}, and the results of including this in our pipeline are shown in blue. This approach dramatically reduces the ILC bias.
 \label{fig:cls_biased_vs_debiased} }
\end{figure}

To use the ILC method, an estimate of the frequency-frequency covariance, $ \mathcal{C}$, is required. We estimate this locally at every point in each of the needlet frames via
\begin{align}\label{eq:ILC_covMat}
\hat{\mathcal{C}}(\mathbf{p})_{ij} = \sum\limits_{\mathbf{p}'} T(\mathbf{p}') m_i(\mathbf{p}+\mathbf{p}')m_j(\mathbf{p}+\mathbf{p}'),
\end{align}
where $T(\mathbf{p})$ is the smoothing kernel. Hereafter we refer to these as ``covariance maps". We choose a smoothed top-hat smoothing function defined as 
\begin{align}
    T(|\bm{\theta}|) = \frac{1}{\left(1+\frac{|\bm{\theta}|}{W_i}\right)^6}.
\end{align} The width, $W_i$, of the top-hat at scale $i$ is set to ensure that we always have a sufficient number of modes so that the covariance matrix is invertible. $W_i$ is computed as
\begin{align}
 W_i = 2\arccos\left(1 - 2 \,\frac{20N^i_\mathrm{elements}}{N^i_\mathrm{modes}} \right),
\end{align}
where $N^i_\mathrm{elements}$ is the number of maps at that needlet scale and $N^i_\mathrm{modes}$ is the number of harmonic modes selected by the needlet spectral funciton, $h^{(i)}(\ell)$. 
 See the Appendix of Ref.~\citep{delabrouille2009} for details of this computation. This top-hat smoothing kernel was used as it was found to produce stable measurements of the covariance matrix, especially near the edges of the maps. 

When the covariance matrix is estimated from the data, it is well known that a bias can be produced in the component-separated maps \citep{Tegmark_2000,Eriksen_2004,Saha_2008,delabrouille2009}. This bias, known as the ILC bias, arises due to chance correlations between the noise and the signal. Modes where the signal and noise cancel have lower variance. These modes are up-weighted in the ILC and this leads to a suppression of the signal. We refer the reader to the appendices of Ref.~\citep{delabrouille2009} for a thorough discussion of this effect. The size of this bias depends on the number of independent modes used to estimate the covariance matrix. When a small number of modes is used, the bias is large. 
For harmonic ILC analyses, where the empirical covariance matrix is estimated via the standard power spectrum, this bias is only significant for the largest scales since at larger $\ell$ there are more $m$ modes to estimate the power spectrum and so a smaller bias.  However, for the needlet ILC presented in this work the problem can affect all scales.  To demonstrate this we simulate a subset of the ACT and \textit{Planck} observations and compare the output ILC to the known input CMB map. For computational speed we simulate a subset of the data: two ACT DR4 maps (at f$150$), two ACT DR6 maps (at f$150$ and f$220$), and three \textit{Planck} maps (at 100, 143 and 217 GHz);
 the results should be similar for the full data set. Fig.~\ref{fig:cls_biased_vs_debiased} shows that the resulting map is biased on all scales. This effect arises from one of the strengths of the needlets: the ability to capture spatial variation across the map. In order to maximally account for spatial variation it is best to estimate the needlet covariance matrix on the smallest possible patch of sky, i.e., using a small smoothing kernel in Eq.~\ref{eq:ILC_covMat}. However, using a small patch of sky means that only a small number of independent modes are used to estimate the covariance matrix. Thus there is a large ILC bias.\footnote{ There is analogous bias in measurements of cluster properties with matched filter methods. There are parallels between our method to mitigate the ILC bias and the method proposed in Ref.~\citep{Zubeldia_2022} to mitigate matched filter biases.}

In the literature there are a range of approaches to minimize this bias: one could use sufficiently large smoothing scales so the bias is small \citep{delabrouille2009}, at the possible cost of a loss of the ability to capture spatial variation; the ILC bias could be approximated analytically or computed via simulations and removed \citep{Basak_2012}; or the ILC method could be modified to minimize a different objective \citep{Remazeilles_2011,Hurier_2013}. In this work we implement a simple alternative method that is motivated by the ILC philosophy of minimal assumptions. Our new approach is to ensure the weights are independent from the data vector by explicitly excluding the data modes, $d_{\nu_i}(\mathbf{x})$, from the calculation of the weights at $\mathbf{x}$, $w(\mathbf{x})$, as detailed  below. This approach never weights any mode by itself and thereby removes the bias.

\begin{figure}
    \centering
  \subfloat[Standard covariance matrix estimation]{\label{fig:schematic_fourier_space}%
  \includegraphics[width=0.4\textwidth]{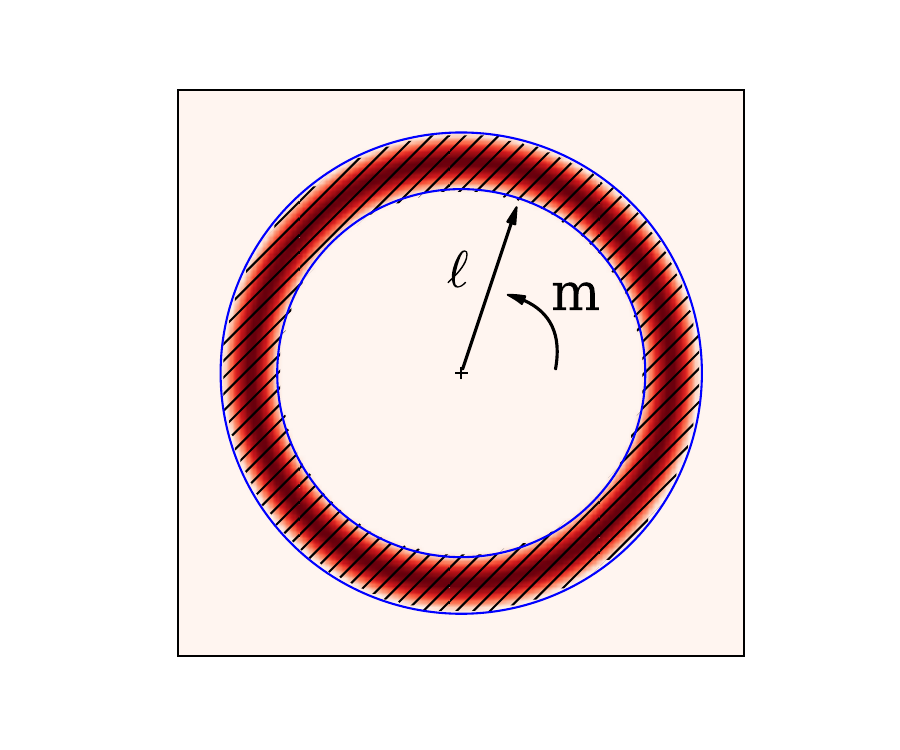}%
 }
 \\
  \subfloat[Harmonic-space bias reduction method]{\label{fig:schematic_mode_isolation}%
  \includegraphics[width=0.4\textwidth]{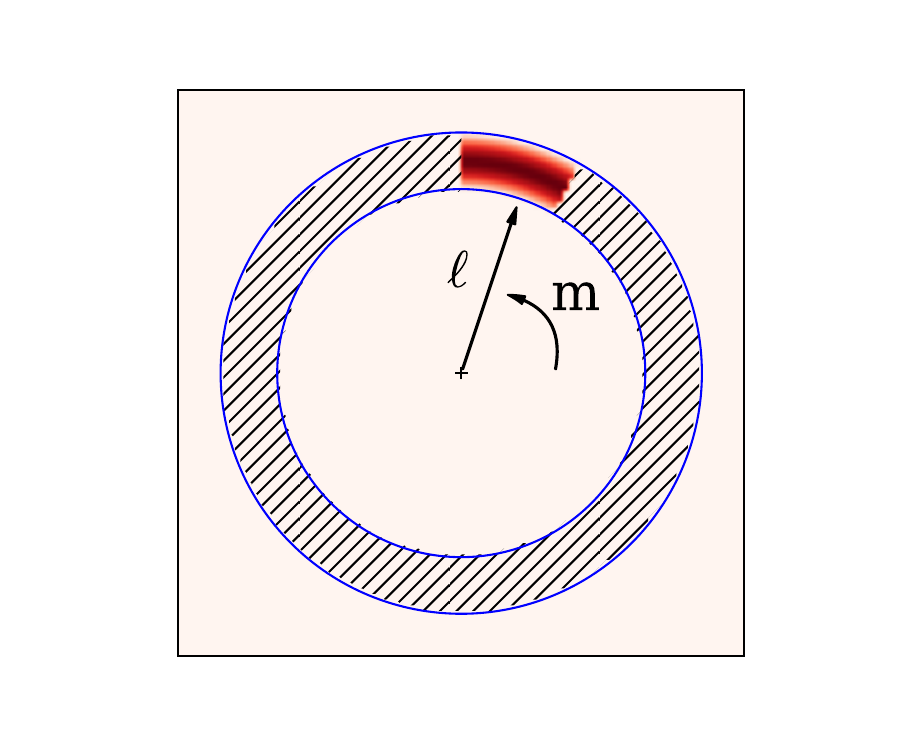}%
 }
 \caption{A schematic demonstrating the large-scale ILC bias mitigation strategy. The region bound by the blue lines denotes the modes selected at one needlet scale in harmonic space. Radial distance corresponds to $\ell$ and azimuth to $m$. The cross-hatched region shows modes used to estimate the weights and the red region shows the modes to which the weights are applied, hereafter the data vector. In the standard method, (a), the weights are computed from all the modes within the needlet band. The double use of each mode, in the weights and the data vector, leads to the ILC bias. In our mitigation method, (b), the weights are computed from the majority of modes within the needlet band, but we explicitly exclude the data vector modes.   \label{fig:LargeScalesBiasRed} }
\end{figure}

\begin{figure}
    \centering
  \subfloat[Standard covariance matrix estimation]{\label{fig:schematic_real_space}%
  \includegraphics[width=0.4\textwidth]{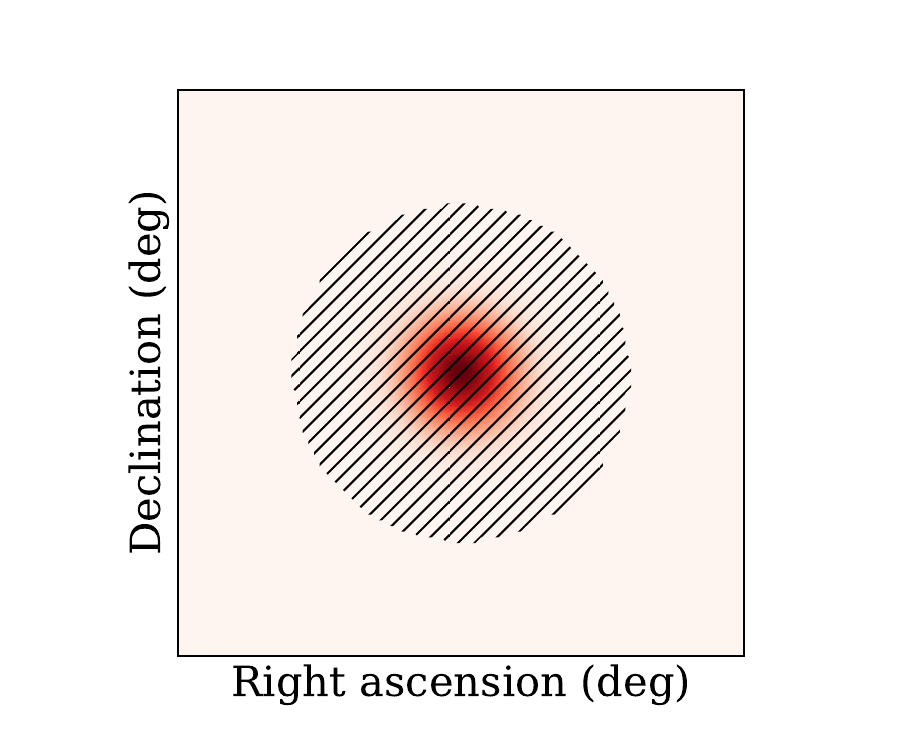}%
 }
 \\
  \subfloat[Real-space bias reduction method]{\label{fig:schematic_donut}%
  \includegraphics[width=0.4\textwidth]{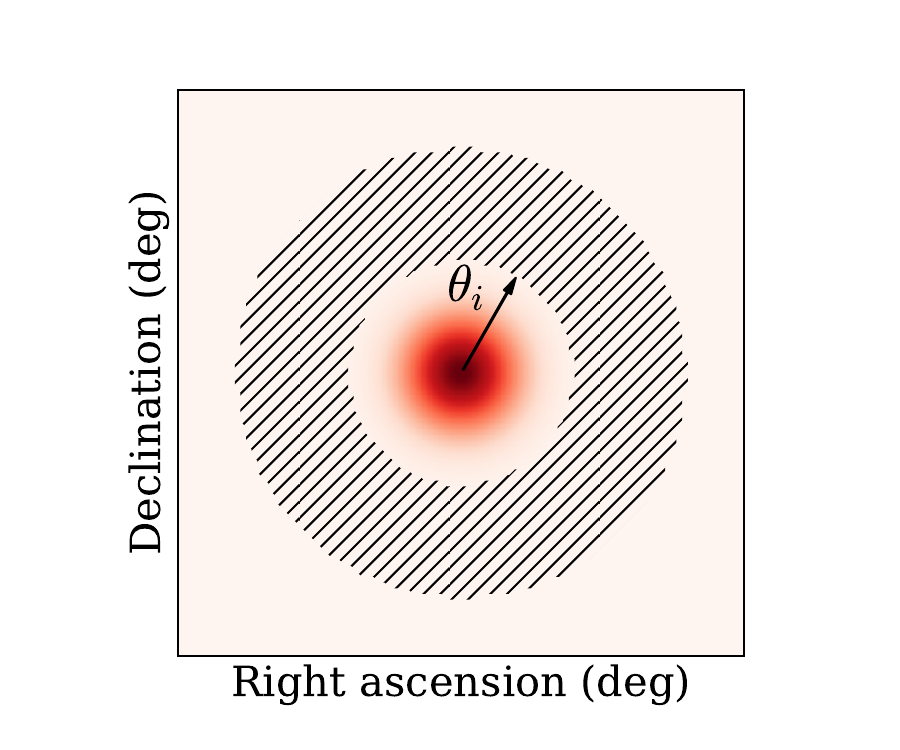}%
 }
 \caption{A schematic demonstrating the small-scale ILC bias mitigation strategy. These plots show the modes selected at one needlet scale in real space. As in Fig.~\ref{fig:LargeScalesBiasRed}, the cross-hatched region shows modes used to estimate the weights and the red region shows the modes to which the weights are applied, that is the data vector. The double use of modes in the standard ILC, (a), for the weights and the data vector, again leads to the ILC bias. In our mitigation strategy, (b), we simply exclude modes within $\theta_i$, the localization scale of the needlet, from the weights. \label{fig:SmallScalesBiasRed} }
\end{figure}
We trivially demonstrate this for the harmonic ILC method. As is detailed in Appendix \ref{app:HarmonicILC}, in the harmonic ILC the weights are applied to the $a_{\ell m}$ modes. The covariance used to compute the weights is the empirical power spectrum. In this case, we can estimate the power spectra using all the $m$ modes except the specific $m$-mode we are trying to reconstruct. This simple modification mitigates the ILC bias. We develop an analogous approach for the needlet ILC; however, isolating subsets of modes is more expensive in the needlet frame. To ensure the method is computationally feasible we use a slightly different approach for the large and small scales -- large scales here correspond to $\ell\lesssim 500$. The approach for the large scales exploits the needlet localization in harmonic space, whilst the small scales exploits the localization in real space.

{\it Large-scale strategy:} This is very similar to the harmonic ILC case. We can remove a small subset of $a_{\ell m}$ modes from the needlet map. The remaining modes can be used to compute a `covariance map', via Eq.~\ref{eq:ILC_covMat}, and a set of ILC weights. These weights are then applied to the small number of modes held out.  As the weights are now uncorrelated with the data, the ILC bias will be removed. This procedure can be repeated on other subsets until weights have been constructed for all the modes. We refer to this as the `harmonic-space bias reduction method'. A schematic of this is shown in Fig.~\ref{fig:LargeScalesBiasRed}. In the needlet frame, this procedure requires four SHTs for each subset considered. As the majority of modes are used in the `covariance map' calculation, we do not need to enlarge the smoothing kernel in Eq.~\ref{eq:ILC_covMat}. In Appendix \ref{app:needlet_m_modes} we provide further details of how the subsets are created. 

{\it Small-scale strategy:} The computational cost of SHTs on small scales motivates us to consider an alternative method for these scales. For these scales we use that fact that the needlets at scale $i$ are spatially localized within some region, $\theta_i$, to construct uncorrelated weights. We use a `donut'-like smoothing kernel to estimate the `covariance maps', given in Eq.~\ref{eq:ILC_covMat}. The hole of the donut is set so that the weights will have minimal contribution from the region within $\theta_i$ but, through the ring around it, still capture the local map properties. This trivial change to the method can be implemented in a computationally efficient manner --- the smoothing is performed with one pair of SHTs. However, it requires slightly larger smoothing scales to ensure there are still enough modes for a stable estimation of the covariance matrix. On the smallest scales we can afford to have slightly larger smoothing scales as the signals tend to vary over scales larger than our smoothing scales. Note we do not use this approach on large scales as we do not want to enlarge the smoothing kernel. We refer to this as the `real-space bias reduction method' and a schematic of this approach is shown in Fig.~\ref{fig:SmallScalesBiasRed}.

In Fig.~\ref{fig:cls_biased_vs_debiased} we show the result of implementing these two methods in our needlet ILC pipeline. These methods reduce the ILC bias by an order of magnitude. The ILC bias is not perfectly removed as neither method produces completely uncorrelated weights: couplings between $m$-modes, induced by effects such as masking, mean that weights from the harmonic-space method are not completely independent from the data modes. For the real-space method, the small overlap between modes in the `donut' and  the data modes leads to slight correlations between the weights and data. 

When computing the covariance matrix we use one further simplification to reduce the computational cost. We use coadded covariance maps to compute the off-diagonal covariance matrix elements. In our data set we have many measurements of the same patch of sky, with approximately similar noise levels but from different detectors. For example we have five maps of the `D56' region in the f150 band. For each such group of maps, we compute an inverse noise variance weighted coadd and use this map for the off-diagonal covariance matrix. This saves computational time because instead of computing the covariance between each map in the group with every other map in the data set, we just need to compute the covariance between the coadded map and the other maps. This assumes that each map within the group has a similar correlation with the remainder of the data set, which is generally very accurate. Note that this does not lead to biases in the ILC map as generally a misestimation of the ILC covariance matrix leads to suboptimal but unbiased ILC maps. Compared to simply using this coadded map as the input to the ILC, this approach has the advantage that we can easily account for differences in the passbands, gains, and frequency-dependent beams of the observations. 

\subsubsection{Frequency Response Functions}\label{sec:freqResponseFunc}
In addition to the covariance matrix, the ILC method requires a response function that characterizes the strength of each of the components of interest at the frequency of each map. The maps used in this work are all converted to ``linearized differential thermodynamic units," i.e., those in which the response to the primary CMB anisotropies is unity. Thus the frequency response function for the CMB and kSZ effect, in Eq.~\ref{eq:basic_ilc_weights}, is simply
\begin{align}
r^\mathrm{CMB}_\nu = 1.
\end{align}

For all other sky signals the response function requires integrating the spectral function, $f_\mathrm{X}(\nu)$, against the instrument passbands, $\tau(\nu)$, as
\begin{align}\label{eq:responseFunc}
r^\mathrm{X}_{\nu,\ell} = \frac{ \int \mathrm{d}\nu f_\mathrm{X}(\nu)\left.\frac{\mathrm{d}B(\nu,T)}{\mathrm{d}T}\right|_{T=T_\mathrm{CMB}}  \nu^{-2}\tau(\nu)}{ \int \mathrm{d}\nu\left.\frac{\mathrm{d}B(\nu,T)}{\mathrm{d}T}\right|_{T=T_\mathrm{CMB}}  \nu^{-2}\tau(\nu)},
\end{align}
where the derivative of the blackbody ${\mathrm{d}B(\nu,T)}/{\mathrm{d}T}$ is the conversion to the ``linearized differential thermodynamic units". The $\nu^{-2}$ factors are there by convention: we assume the passband $\tau(\nu)$ quantifies the response to a Rayleigh-Jeans ($\nu^2$) source. 

In general, the spectral function for the tSZ effect depends weakly on the temperature of the electrons, $T_e$, that scatter the CMB photons \citep{Challinor_1998,Itoh_1998}. Given that the temperature of electrons varies throughout the universe, no single temperature value will capture all of the tSZ signal. In this work we consider two approaches: 1) to neglect the temperature-dependent terms, as done in most previous component separation analyses, and 2) to use the scale-dependent temperature proposed in Ref.~\citep{Remazeilles_2019}. The assumption that the temperature dependence can be ignored is commonly used in the literature \citep[see e.g.][]{planck2016-l04,Madhavacheril_2020,Bleem_2022}, captures most of the signal, and has a simple analytic form. The second case provides a more accurate extraction of the tSZ anisotropies, at the cost of a more complex analysis.    

The temperature-dependent terms are important when the electrons are relativistic, i.e., $k_B T_e\sim m_e c^2$. The assumption that we can ignore these terms is justified as most electrons in the universe are non-relativistic, i.e., have temperatures $T_e\ll m_e c^2/k_B $ where $m_e$ is the electron mass, $c$ is the speed of light, and $k_B$ is Boltzmann's constant. In the non-relativistic regime, the tSZ frequency response function is independent of the electron temperature and has the following analytic form:
\begin{align}\label{eq:tSZResponse}
f_\mathrm{tSZ}(\nu) = \frac{h \nu}{k_B T_\mathrm{CMB}} \frac{\exp\left[ \frac{h \nu}{k_B T_\mathrm{CMB}}\right] + 1}{\exp\left[ \frac{h \nu}{k_B T_\mathrm{CMB}} \right] -1} - 4,
\end{align}
where  $h$ is Planck's constant and $T_\mathrm{CMB} = 2.726\,$K is the temperature of the CMB \citep{fixsen1997,fixsen2009}. In keeping with past work, this is our baseline choice for the tSZ spectrum.

\begin{figure}
    \centering
  \includegraphics[width=0.48\textwidth]{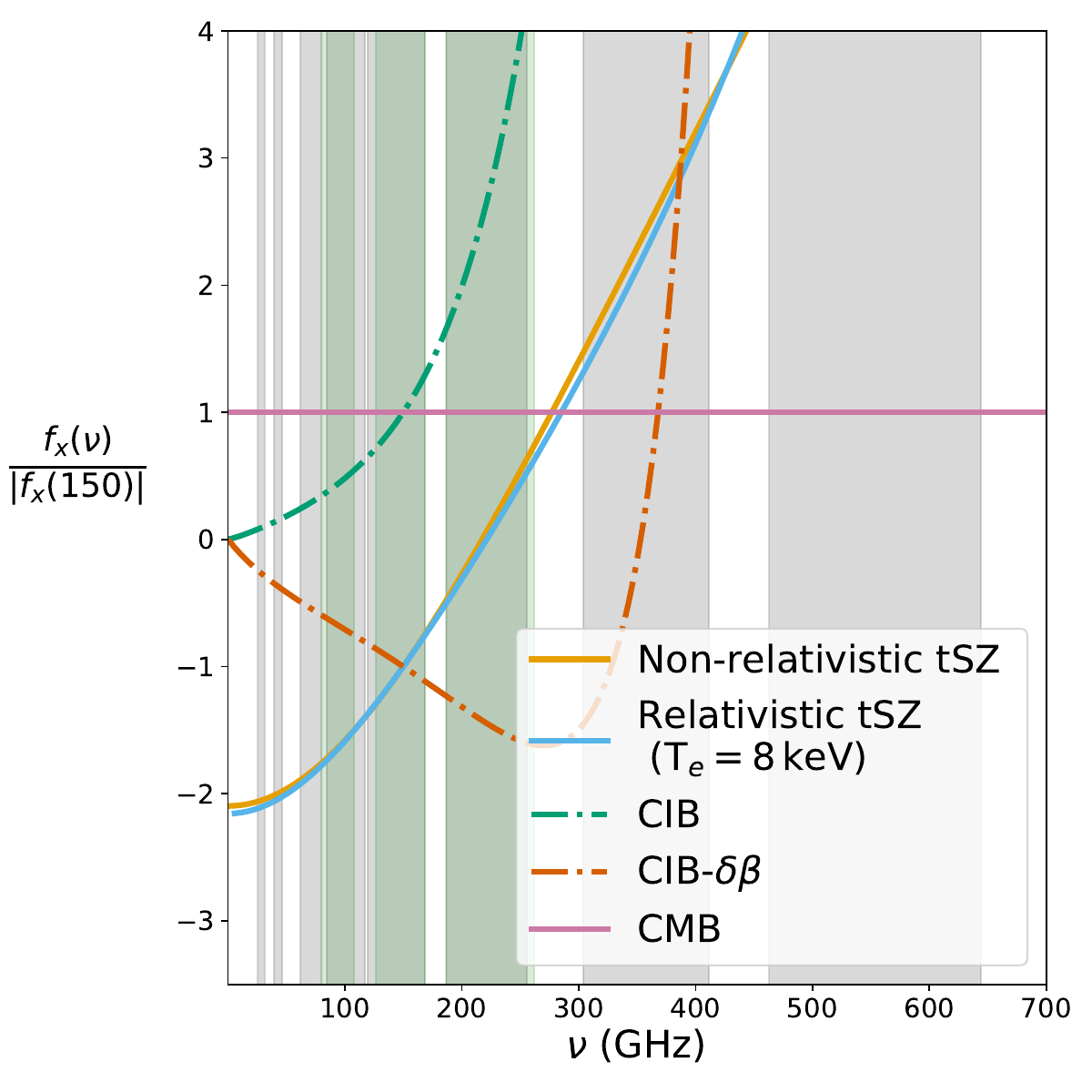}%
 \caption{ The frequency dependence of the sky signals considered in this work, normalized by the magnitude of the response at 150 GHz. The CMB is shown in pink. The tSZ frequency response depends upon the temperature of the electrons, $T_e$, that scatter the CMB photons. Here we compare two cases: first when the temperature is sufficiently low that the electrons are non-relativistic, (gold), and second when the temperature is $T_e=8$\,keV (blue). In addition to producing a minimum-variance map of the signal of interest, it is often necessary to ensure this map is not contaminated with other sky signals, especially the CIB. To do this we explicitly remove the CIB using the response shown in dot-dashed green. To account for differences between the model CIB spectral energy distribution (SED) and the true SED, we remove sky signals consistent with a second SED, denoted CIB$-\delta \beta$. This second template is obtained from a Taylor expansion of the model SED around the fiducial spectral index, $\beta$. In gray and light green we show a measure of the \textit{Planck} and ACT passbands respectively. Note that the responses are in ``linearized differential thermodynamic'' units, hence the unit response of the CMB at all frequencies.
 \label{fig:spectral_responses} }
\end{figure}

For the second approach, we use \textsc{szpack} \citep{Chluba_2012,Chluba_2013} to compute the full tSZ frequency response, $g(\nu,T_e)$.  We refer to the response in Eq.~\ref{eq:tSZResponse} as the non-relativistic response and to the response incorporating temperature dependence as the relativistic tSZ response. In Fig.~\ref{fig:spectral_responses} we compare the non-relativistic frequency function to the relativistic one at $T_e=8$\,keV. Whilst the differences are generally small, they can be important at the precision obtainable with current data.

Following Ref.~\citep{Remazeilles_2019}, we use a different temperature for each $\ell$, denoted as $\bar{T}_{e}(\ell)$. The scale-dependent temperature accounts for the fact that the largest scale tSZ anisotropies come from the most massive and hottest objects, whilst smaller scale anisotropies come from less massive and cooler objects.  The temperature is computed as a Compton-$y$ weighted expectation of galaxy cluster temperatures and we refer the reader to Ref.~\citep{Remazeilles_2019} for more details. The relativistic tSZ response is then 
\begin{align}\label{eq:resp_tsz_scale_dep}
f^\mathrm{tSZ-relativistic}(\nu, \ell) = g(\nu,\bar{T}_{e}(\ell)).
\end{align}
The expected temperatures range from $\sim 8\,$keV on large angular scales to $\sim 2\,$keV on small scales. We discuss the implications of this more extensively in Section \ref{sec:PropComptonY}.

One of the main contaminants for tSZ studies is the cosmic infrared background (CIB). We model the CIB frequency function as a modified blackbody,
\begin{align}
 f_\mathrm{CIB}(\nu) = \frac{A \left(\frac{\nu}{\nu_0}\right)^{3+\beta}}{\exp\frac{h\nu}{k_B T_\mathrm{CIB}}-1} \left(\left.\frac{\mathrm{d}B(\nu,T)}{\mathrm{d}T}\right|_{T=T_\mathrm{CMB}} \right)^{-1},
 \end{align}
where $\beta=1.7$ and $T_\mathrm{CIB}=10.70$\,K are the parameters characterizing the approximate all-sky CIB modified blackbody SED, $\nu_0$ is an arbitrary normalization frequency and $A$ is a normalization constant. These are obtained to fits of a theoretically-calculated CIB monopole at {217, 353,  545} GHz, as calculated in~\citep{planck2013-pip56} using a halo model fit to observations of the CIB anisotropies \citep{McCarthy_2023}.\footnote{Over the range of frequencies probed $\beta$ and $T_\mathrm{CIB}$ are fairly degenerate. This means that the SED used here is consistent with that used in Ref.~\citep{Madhavacheril_2020}, despite the different values of $\beta$ and $T_\mathrm{CIB}$. Ref.~\citep{Madhavacheril_2020} fix $T_\mathrm{CIB}$ to a higher value $T_\mathrm{CIB}=24$\,K, which leads to a lower inferred value of $\beta$. }

Completely removing the CIB is challenging for two reasons: first, the spectral function is not well known; the functional form used above is a theoretically motivated empirical fit. Deprojecting an inaccurate template leaves residual CIB. Second, the CIB anisotropies exhibit decorrelation across frequencies --- i.e., the anisotropies at one frequency are not perfectly correlated with those at a second \citep{planck2013-pip56,Lenz_2019}. This behavior can be understood with a toy model: consider a case where the rest frame SED for the CIB galaxies is the same for all sources. The observed emission is then given by the sum of the redshifted emission from sources over a wide redshift range. This redshifting means that the observed emission at a single frequency comes from many different source frequencies, or equivalently the observed SED of each galaxy is different. Galaxies at different redshifts will then provide a different relative contribution at different observational frequencies. %
As different frequencies probe different redshift weightings of the sources, the spatial anisotropies will not be perfectly correlated. 

For many applications having a small level of residual CIB will not produce any biases; however, for some applications this can be critical. To mitigate the residual CIB we use the method of Ref.~\citep{Chluba_2017}, hereafter referred to as the moment expansion method --- we deproject a second spectral template that represents a Taylor expansion about the assumed CIB spectral index, $\beta$. This model was shown to better account for SEDs that are a sum of modified blackbody spectra, and consequently is a better approximation of the CIB \citep{Chluba_2017}. For such studies we provide maps that additionally deproject the derivative spectrum given by
\begin{align}\label{eq:derivBeta}
 f_\mathrm{CIB-\delta \beta}(\nu) = \frac{A \ln(\nu/\nu_0) \left(\frac{\nu}{\nu_0}\right)^{3+\beta}}{\exp\frac{h\nu}{k_B T_\mathrm{CIB}}-1} \left(\left.\frac{\mathrm{d}B(\nu,T)}{\mathrm{d}T}\right|_{T=T_\mathrm{CMB}} \right)^{-1}.
 \end{align}
Note that each additional deprojection comes at an additional noise cost in the final ILC map, as the additional constraints lead to a less-minimal-variance solution.

To compute the responses with Eq.\ref{eq:responseFunc}, we need the instrument passbands. For ACT we use the Fourier transform spectrometer measurements reported in Ref.~\citep{Thornton_2016} for the DR4 data and an upcoming paper for the equivalent DR6 data. For the \textit{Planck} passbands we use those from \cite{zonca2009,planck2013-p02,planck2013-p03d}, and we additionally include central frequency shifts as reported in Ref.~\citep{planck2014-a12} and Ref.~\citep{planck2020-LVII}.

As is discussed in Appendix A of Ref.~\citep{Madhavacheril_2020}, the finite width of the passbands means that there is a different effective beam for each of the components on the sky. This arises from the combination of two effects: first, each sky signal has a different frequency dependence and thus is more important in different parts of the instrumental passbands. Second, the beam is different at different frequencies (for diffraction-limited optics we have FWHM $\propto 1/\nu$). In this work we follow the method of Ref.~\citep{Madhavacheril_2020} and account for this using scale dependent ``color corrections''.\footnote{We note that the scale-dependent ``color corrections'' discussed here are related to, but distinct from, the ``color corrections'' described in, e.g.,Ref. \citep{Griffin_2013} and Ref. \citep{planck2013-p03d} This will be expanded upon in a forthcoming paper.} Scale-dependent color corrections are changes to the frequency response functions that account for the different effective beam seen by each signal. At large scales this effect is negligible; however, the color corrections can be $\gtrsim 10\%$ changes on small scales.

First we compute the scale-dependent responses as
\begin{align}\label{eq:scaleDepResp}
r^{X}_{\nu,\ell} = \frac{\int\mathrm{d}{\nu} b(\ell,\nu)f^X(\nu) \left.\frac{\mathrm{d}B(\nu,T)}{\mathrm{d}T}\right|_{T=T_\mathrm{CMB}} \nu^{-2}\tau(\nu) }{b(\ell,\nu_0)\int\mathrm{d}{\nu}\left.\frac{\mathrm{d}B(\nu,T)}{\mathrm{d}T}\right|_{T=T_\mathrm{CMB}} \nu^{-2}\tau(\nu)},
\end{align}
where $b(\ell,\nu)$ is the frequency dependent beam and $\nu_0$ is the reference frequency. The details of the computation of the frequency-dependent beams will be provided in an upcoming ACT paper. Then we compute the weighted average of this across the needlet band to get each component's response in that band as
\begin{align}
r^{(i),X}_{\nu} = \frac{\sum\limits_\ell r^{X}_{\nu,\ell}h^{(i)}_{\ell}}{\sum \limits_\ell h^{(i)}_{\ell}}.
\end{align}
Again following Ref.~\citep{Madhavacheril_2020} we do not apply scale-dependent corrections to the lower resolution \textit{Planck} data as the investigations in Ref.~\citep{planck2013-p02,planck2013-p03d} found no evidence for scale-dependent color corrections.

\begin{figure}
    \centering
  \includegraphics[width=0.48\textwidth]{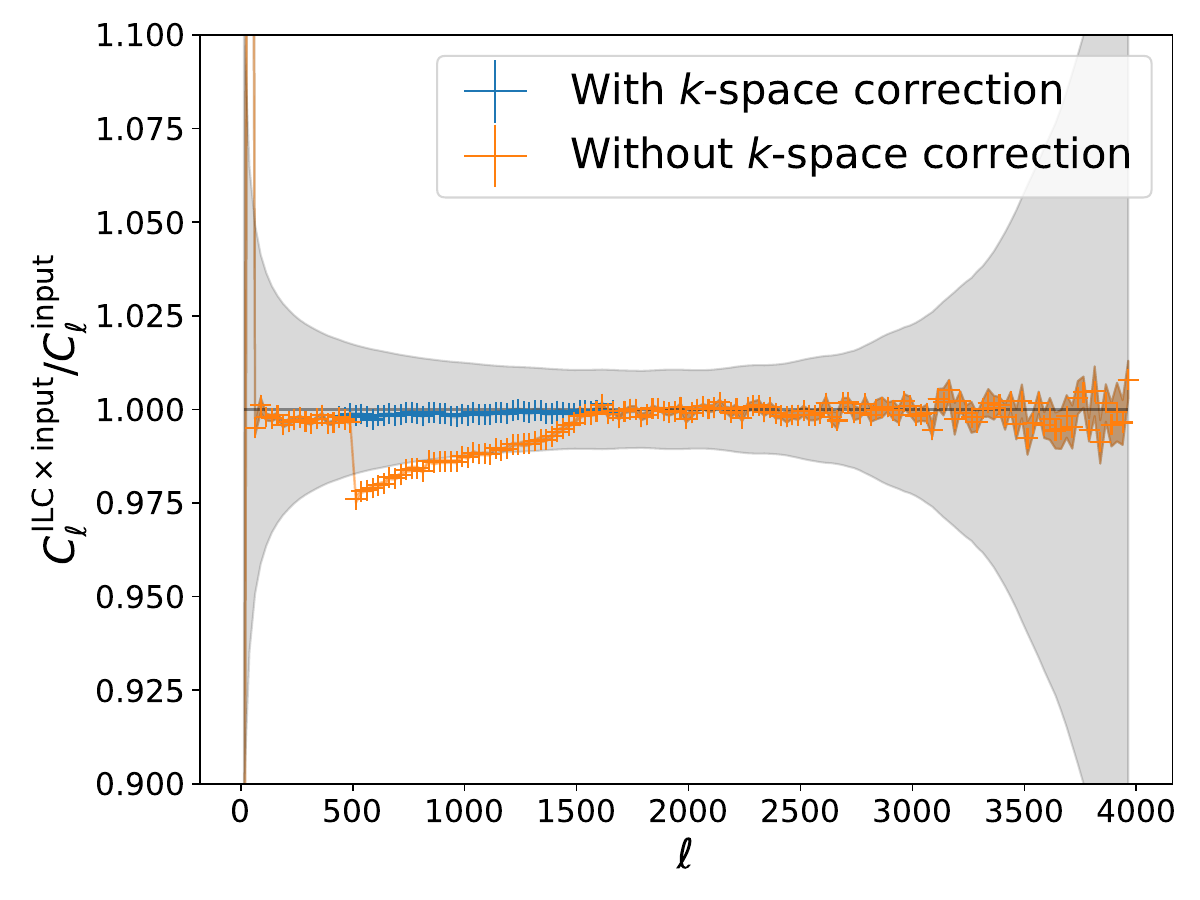}%
 \caption{The ACT data are filtered to remove a set of bright contaminants, such as scan-synchronous pickup --- see Section \ref{sec:filteringCorrection} for more details.  The orange points show the ratio of the cross-correlation of the ILC map, including the final filter, and input component map to the input map auto-power spectrum. This demonstrates that a non trivial set of modes are impacted by the Fourier-space filter. Using the correction method described in Section \ref{sec:filteringCorrection} we can correct for these missing modes, with the result shown in blue. With the correction there are no significant biases induced by the NILC pipeline. For reference the gray band shows the $1\sigma$ error on the ILC output power spectrum. The blue and orange error bars are the error on the mean. These errors are very small on large scales as the sample variance cancels in the ratio. For this analysis we use Gaussian simulations of the complete ACT \& \textit{Planck} data set. The largest scales, $\ell\lesssim500$, only use \textit{Planck} data, which does not need Fourier-space filtering. The Fourier-space correction thus only impacts larger $\ell$, hence the feature seen at $\ell \approx 500$.
 \label{fig:cls_with_vs_without_kSpaceCorrection} }
\end{figure}

\subsection{Inverse Needlet Decomposition}
The ILC method produces a set of component-separated maps at each needlet scale, $\hat{s}^{(i)}(\mathbf{p})$. We then need to recombine the maps into a single real-space map, $\hat{s}(\mathbf{n})$. One of the key features of needlets is that this operation is straightforward: one simply convolves each needlet map with the needlet kernel associated with that scale and sums all the resulting maps, i.e.,
\begin{align}
\hat{s}(\mathbf{n}) = \sum\limits_{\ell m} Y_{\ell m}(\mathbf{n})  \sum\limits_i h^{(i)}_{\ell} \sum\limits_{\mathbf{p}} w(\mathbf{p}) Y^*_{\ell m}(\mathbf{p})\hat{s}^{(i)}(\mathbf{p}).
\end{align}
Note that for this operation to not lose any information over the scales of interest we require that the needlet spectral functions satisfy 
\begin{align}
\sum_i \left(h^{(i)}_{\ell}\right)^2=1.
\end{align}
\begin{figure*}
    \centering
  \subfloat[ACT \& \textit{Planck} NILC CMB temperature and kinetic Sunyaev-Zel'dovich anisotropy map.]{\label{fig:full_sky_tmap.}%
  \includegraphics[width=0.99\textwidth]{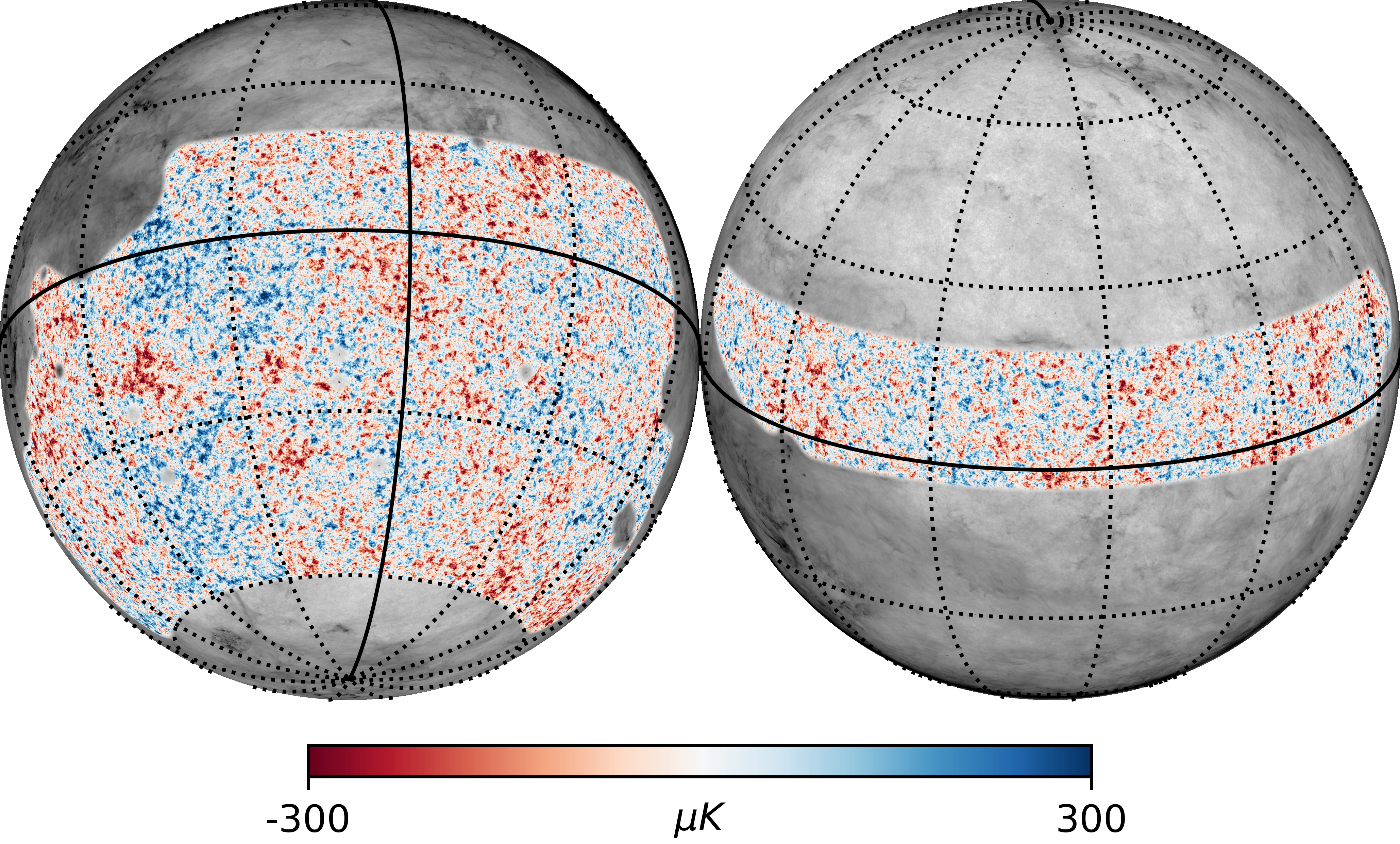}%
 }
 \\
  \subfloat[ACT \& \textit{Planck} NILC CMB E-mode anisotropy map.]{\label{fig:full_sky_emap}%
  \includegraphics[width=0.95\textwidth]{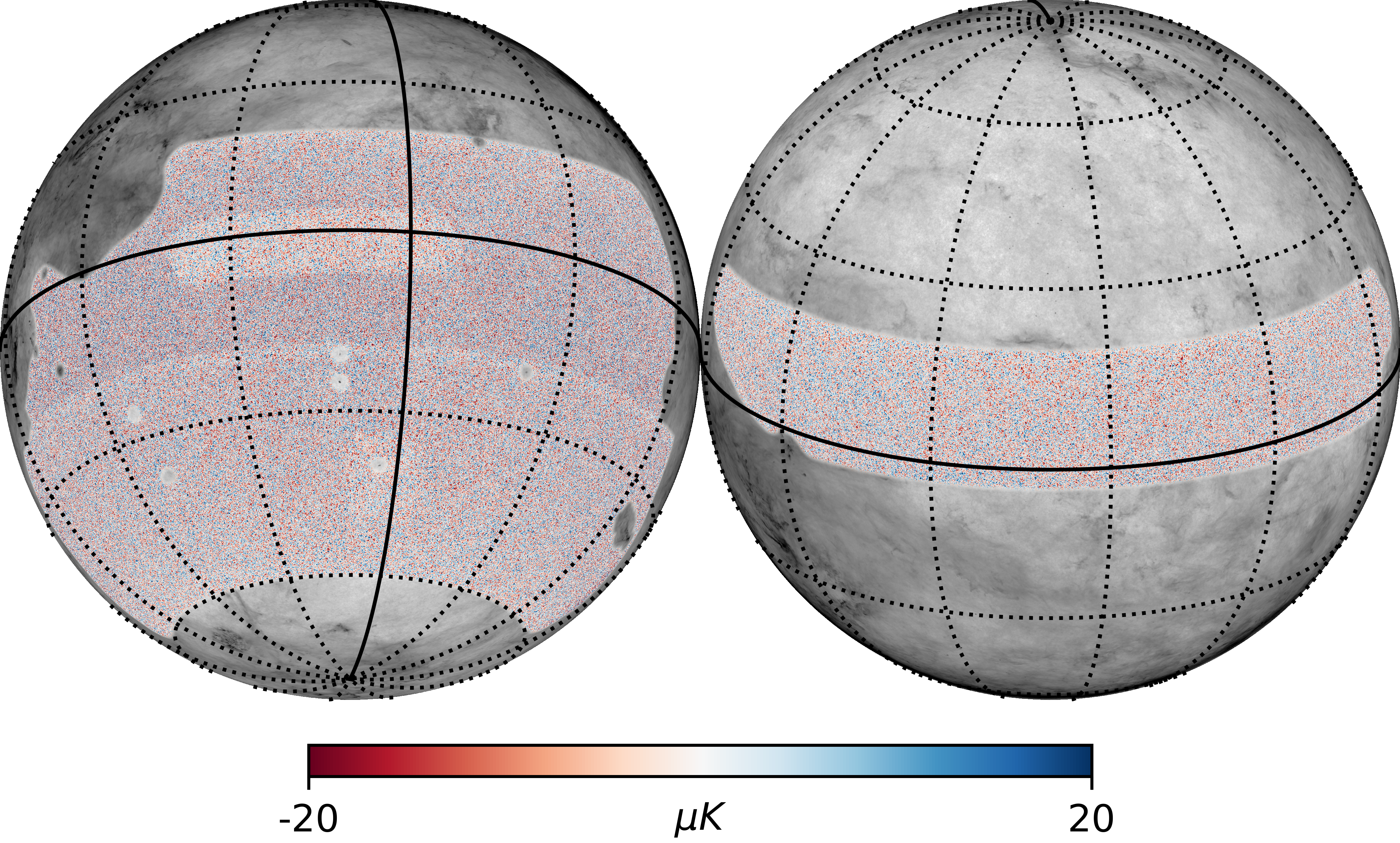}%
 }
  \caption{Maps of two sky components across the full footprint used in this work. Within each of the CMB temperature (a) and CMB E-mode (b) anisotropy maps we can see large-scale CMB fluctuations. The variations in the ACT depth are visible as changes in the small scale noise in the E-mode map. Note that to aid visualization the color scale saturates. The gray scale image in the background is \textit{Planck} Commander dust map. Bright dusty star forming galaxies and clumps of galactic dust can be seen in many of the holes of the mask, see Fig.~\ref{fig:mask_footprints}.\label{fig:FullSkyMaps} }
 \end{figure*}%
\begin{figure*}[ht]
  \subfloat[ACT \& \textit{Planck} NILC Compton-$y$ map.]{\label{fig:full_sky_ymap}%
  \includegraphics[width=0.95\textwidth]{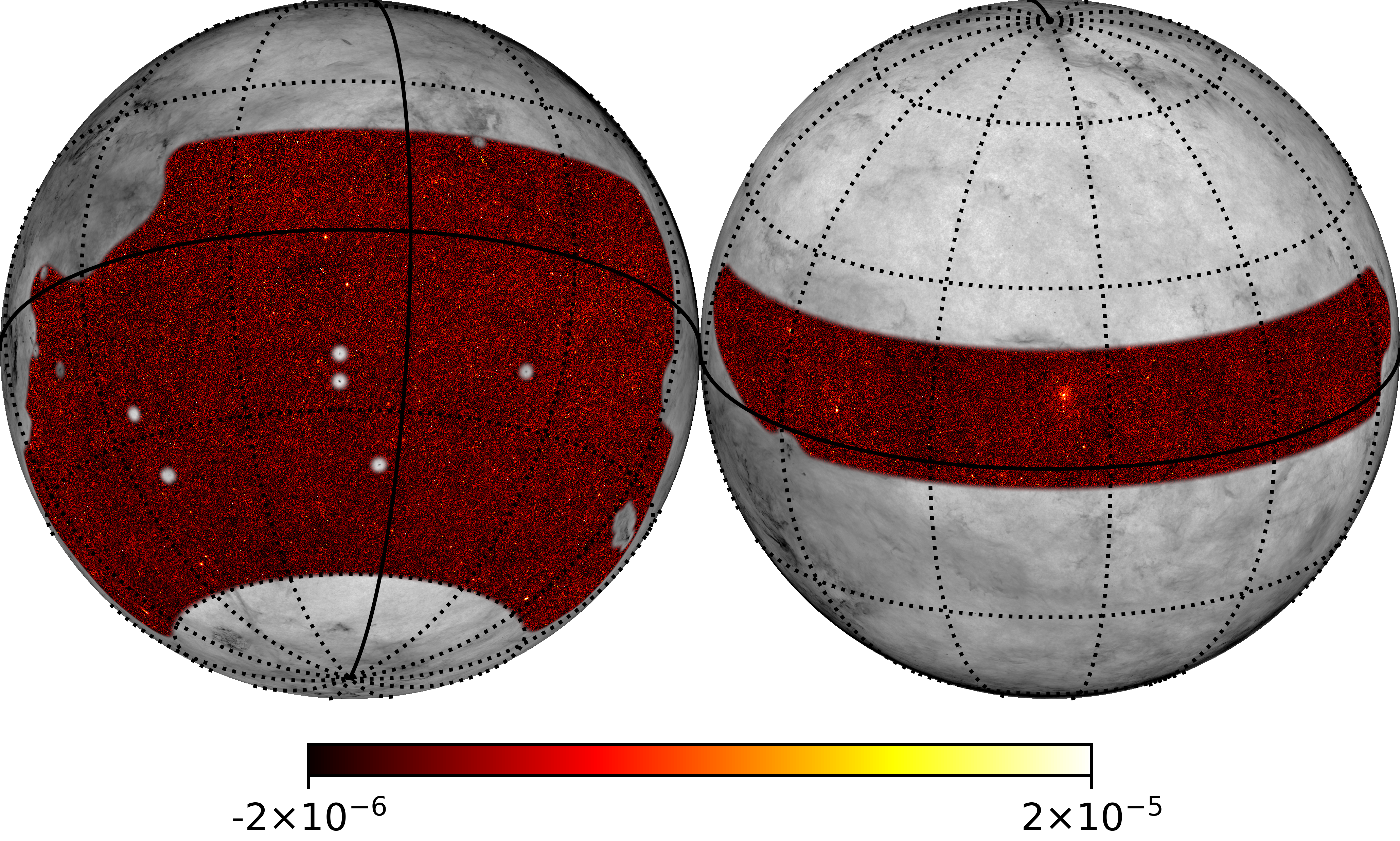}%
 }
 \caption{A map of Compton-$y$ across the full footprint used in this work. We can clearly see many galaxy clusters. As in Fig.~\ref{fig:FullSkyMaps}, to aid visualization the color scale saturates and the gray scale background image is the \textit{Planck} Commander dust map.\label{fig:FullSkyYMap} }
\end{figure*}

\begin{figure*}
    \centering
  \subfloat[CMB temperature and kinetic Sunyaev-Zel'dovich anisotropy map.]{\label{fig:zoom_in_tmap_nokSpaceCorr}%
  \includegraphics[width=0.95\textwidth]{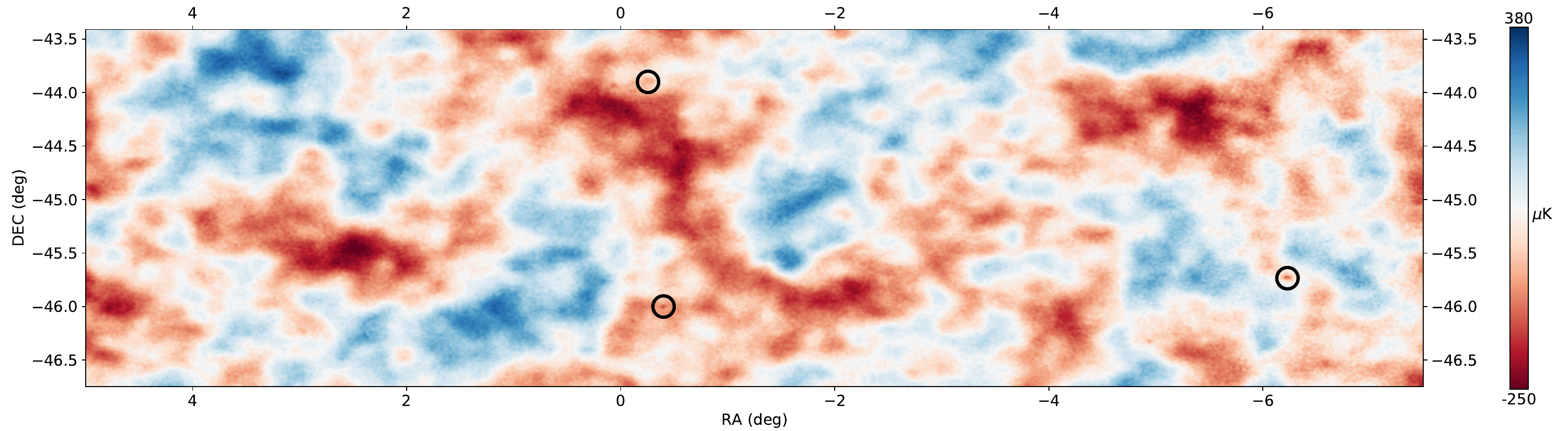}%
 }
 \\ 
   \subfloat[CMB E-mode polarization map.]{\label{fig:e_maps}%
  \includegraphics[width=0.95\textwidth]{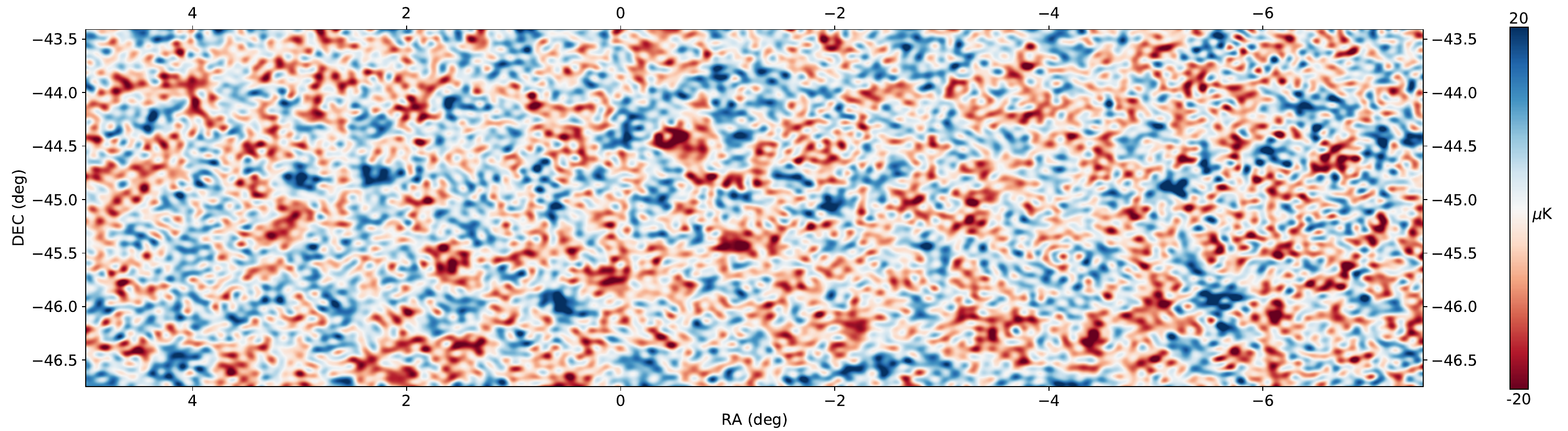}%
 }
 \\ 
  \subfloat[Compton-$y$ map.] {\label{fig:zoom_in_ymap_withkSpaceCorr}%
  \includegraphics[width=0.95\textwidth]{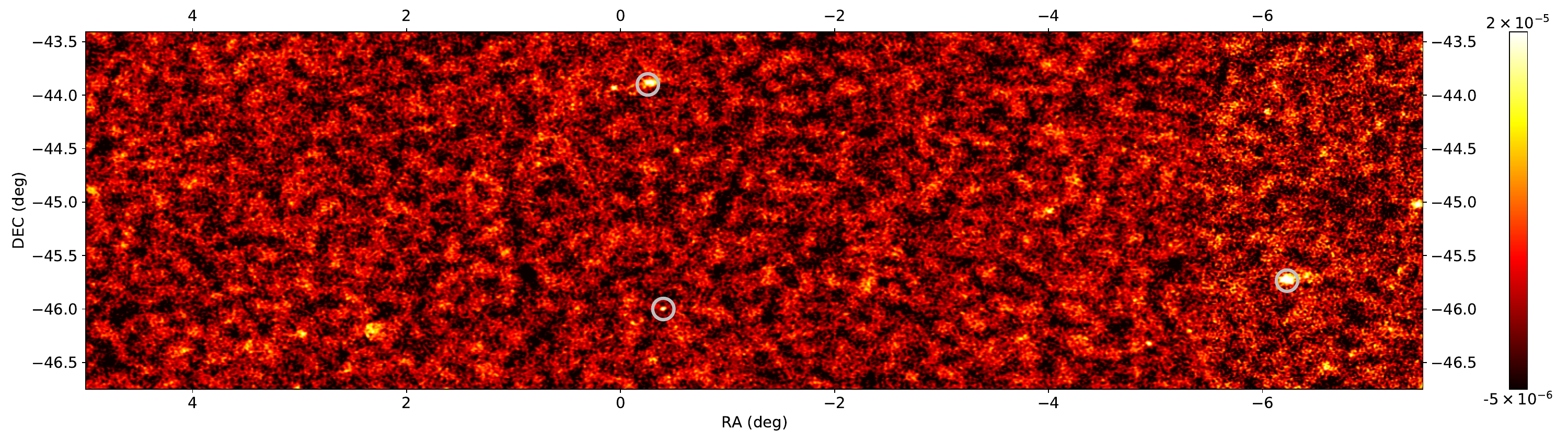}%
 }
 \caption{A $\sim 30$\,deg$^2$ region of the component-separated maps for the CMB temperature, CMB E-mode polarization and Compton-$y$ anisotropies. In (a) and (b) the CMB temperature and E-mode anisotropies are clearly visible across a broad range of scales, whilst in (c) we can see numerous bright clusters picked out by the thermal Sunyaev-Zel'dovich effect. A comparison between the CMB temperature and Compton-$y$ maps shows that the separation of the two signals is not perfect --- small imprints of tSZ clusters can be seen in the CMB temperature map and we highlight three examples with black circles and their corresponding clusters in the Compton-$y$ map. The increase in noise seen at the right edge of the Compton-$y$ map occurs due to changes in the depth of the ACT maps. Note that the maps appear slightly horizontally stretched as the CAR projection is conformal only at the equator.\label{fig:T_and_y_maps}}
\end{figure*}

\begin{figure}
    \centering
  \subfloat[CMB Temperature Anisotropies] {\label{fig:cl_tt} \includegraphics[width=0.45\textwidth]{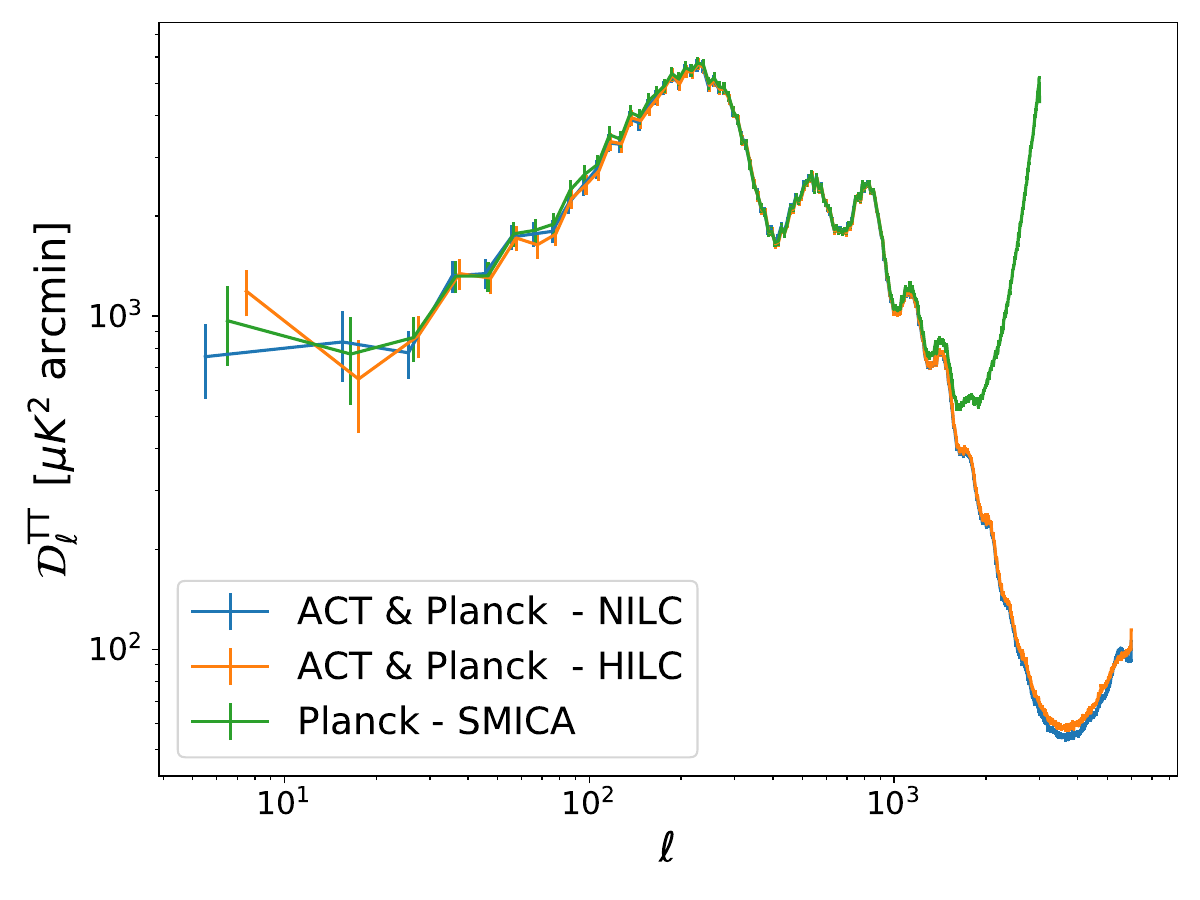} }%
  \\
   \subfloat[Compton-$y$] {\label{fig:cl_yy} \includegraphics[width=0.47\textwidth]{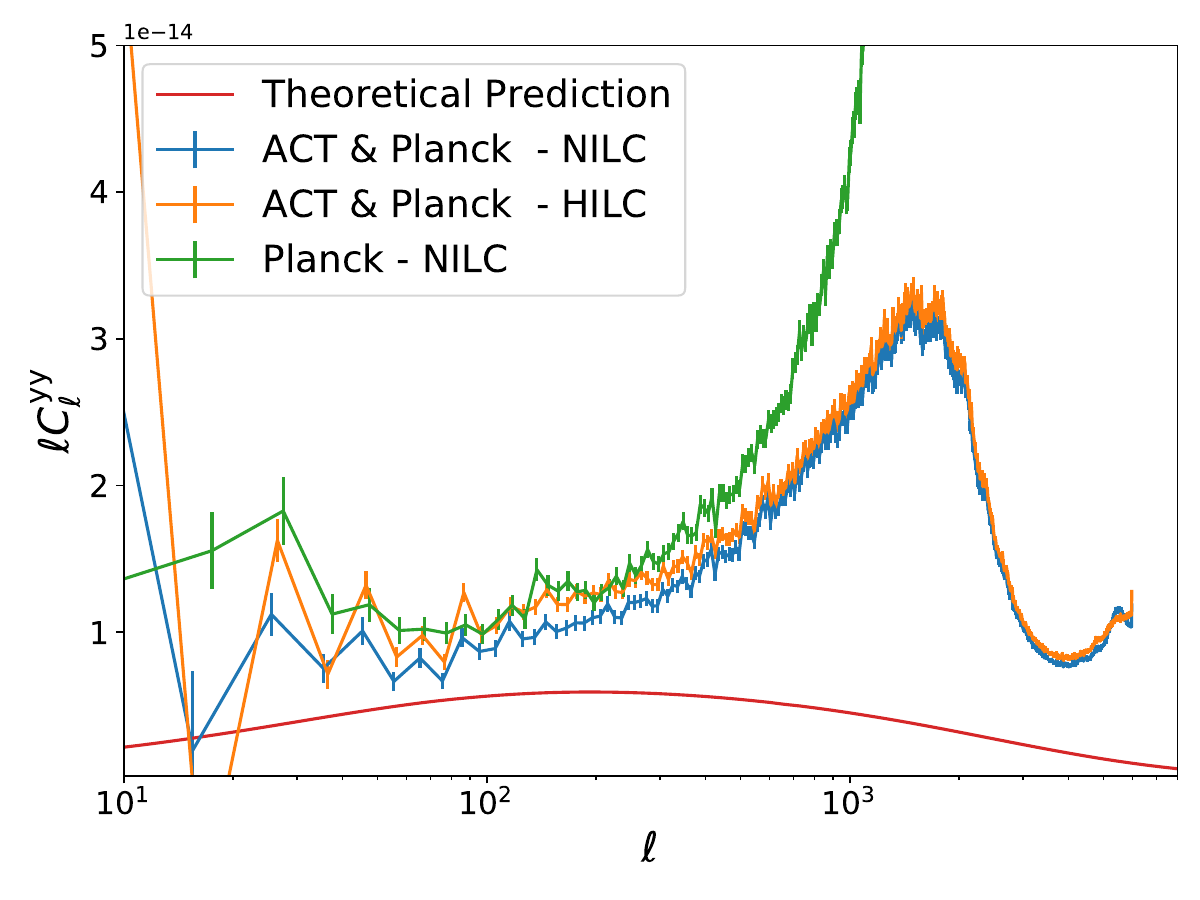}}%
 \caption{The power spectrum of the CMB Temperature and Compton-$y$ NILC maps computed with \textsc{namaster}. We compare these with the corresponding \textit{Planck} \textsc{smica} CMB and NILC Compton-$y$ maps, and harmonic ILC (HILC) maps produced from the ACT and \textit{Planck} data. The large increase in power at high $\ell$ (small scales) in the \textit{Planck}-only data is due to noise in the \textit{Planck} IlC maps, and is significantly reduced in the ACT \& \textit{Planck} ILC maps. The theoretical prediction for the Compton-$y$ power spectrum was computed with \textsc{class-sz} \citep{Bolliet_2022}. The Compton-$y$ map is noise dominated at all scales, hence all the data points lie above the theoretical prediction. The vertical bars denote the errors and are computed analytically via the Gaussian approximation.
 To aid comparisons we have horizontally offset the points in the plot.\label{fig:map_power_spectra} }
\end{figure}
\subsection{Fourier-space Filtering Correction}\label{sec:filteringCorrection}

The Fourier-space filtering performed on the ACT maps corresponds to a highly complex operation in harmonic space. This is especially true once the maps have been combined in the needlet domain with satellite data, for which this filtering is not performed. In the ILC, the filtered modes in the satellite data are treated as `noise', due to their absence in the ACT maps, and so are partially removed. For applications that are driven by the smallest scales, such as cluster stacking investigations, the effect of this filtering is minimal as the filter only removes large scale modes. Thus, for studies on these scales the filtering can likely be safely ignored. However, for applications that require modes with $\ell \lesssim 1500$ this effect is non-trivial and extends beyond the power spectrum in a complex and anisotropic manner.

To avoid having to model this complex, anisotropic effect in future analyses we attempt to correct for it at the map level. This is done by infilling the filtered modes using the \textit{Planck} observations. Specifically, we generate two versions of the component-separated maps: one that uses the complete data set (map A) and another that only uses \textit{Planck} data (map B). We then filter map A with a filter with the same form as the initial filtering, Eq.~\ref{eq:kspaceFilter}, but with the filtering parameters increased ($k_x$ is doubled whilst $k_y^\mathrm{central}$ is increased to 1450).  We then apply the inverse of the Fourier space filter to map B to isolate the removed modes. Finally we add these isolated modes to the filtered version of map A to obtain a corrected map. The purpose of refiltering with an enlarged filter size is twofold: 1) to ensure we better remove residual scan-synchronous pickup; and 2) to provide a well defined set of filtered modes. The ILC does not remove all the filtered modes from the \textit{Planck} data set so the output, uncorrected maps have a complicated effective filtering. Applying the larger filter simplifies the effective filter. Even without the correction described here, such a filter would likely still be needed to ensure that the effective filter is well characterized. In Fig.~\ref{fig:cls_with_vs_without_kSpaceCorrection} we see that this procedure corrects the leading order effect of the Fourier space filtering. Note that the Fourier-space filter only effects modes with $\ell\gtrsim 450$ as below this only \textit{Planck} data is used, for which Fourier-space filtering is not needed.  This method has a cost: the modes added back into the map typically have larger noise than the other modes. This is because they are obtained from \textit{Planck} maps that have lower resolution and higher noise. Whilst these maps thus have anisotropic noise, this is a small effect and can be ignored for most applications. We again note that the filtering used in this work is less aggressive than that used in other ACT analyses, e.g., Refs. \citep{Choi_2020} and \citep{Qu_2022}, and thus the maps will contain some residual scan-synchronous pickup. For cross-correlations this is unimportant, however for analyses using only these maps, such as CMB-lensing reconstruction or primordial non-Gaussianity searches, tests should be performed to see whether this residual impacts the results. If needed, the output ILC maps can be filtered again to remove any residual contamination, with the cost of also removing some signal modes. 

Fig.~\ref{fig:cls_with_vs_without_kSpaceCorrection} demonstrates that with our $k$-space correction and ILC bias reduction methods our NILC pipeline produces unbiased maps. 

\section{Component-Separated Maps}\label{sec:maps}

Using this pipeline we produce the key results of this work: component-separated CMB temperature and E-mode polarization anisotropies and Compton-$y$ maps with a maximum scale of $\ell=17000$ ($\sim 0.6$ arcmin). The full-area maps are shown in Fig.~\ref{fig:FullSkyMaps} and Fig.~\ref{fig:FullSkyYMap}.

In Fig.~\ref{fig:T_and_y_maps} we show a cut out of the maps of CMB temperature, E-mode and Compton-$y$ anisotropies. There are several interesting features: first, through combining the \textit{Planck} and ACT measurements, we obtain a CMB temperature map that is dominated by CMB modes, rather than instrumental and atmospheric noise, over a wide range of scales. Without \textit{Planck} we would not be able to resolve the largest-scale modes and without ACT we would have limited small-scale sensitivity. Secondly, whilst the power spectrum of the Compton-$y$ map is noise-dominated on almost all scales (as discussed in Section \ref{sec:PropComptonY}), we can clearly see many bright compact objects. These objects are galaxy clusters as the Compton-$y$ anisotropies map out the integrated electron pressure,
\begin{align}
    y(\mathbf{n}) = \frac{\sigma_T}{m_ec^2}\int\limits_{los} \mathrm{d}l\,P_e(\mathbf{n},l),
\end{align}
where $\sigma_T$ is the Thomson cross section, $m_e$ is the electron mass, $c$ is the speed of light, and $P_e$ is the thermal electron pressure.
They are visible above the noise due to the highly non-Gaussian structure of the Compton-$y$ signal. The number of these visible by eye is significantly more than can be seen in the individual frequency maps. Thirdly, by comparing the CMB temperature map and the Compton-$y$ map we can see that the separation of the components is not perfect --- the imprint of the brightest SZ clusters can still be seen in the CMB temperature map. As was discussed in Section \ref{sec:ILC_intro}, this arises as the standard ILC method minimizes the total `noise' and not the contribution of any individual foreground sky components.  For some analyses this residual extragalactic contamination can be problematic. To alleviate this, we provide CMB temperature maps that are explicitly constructed to have no contribution from the tSZ signal. This comes at the cost of slightly higher noise. These maps  are further discussed in Section \ref{sec:CMB_map_properties}.

The next notable feature can be seen around right ascension (RA) $\sim -5.5^\circ$, where the noise to the left of this line is noticeably lower. The region shown in this cut out coincides with the edge of the ACT ``D8" field, as is seen in Fig.~\ref{fig:mask_footprints}, and the lower noise occurs through the inclusion of this deep observation in the output map.  This boundary highlights a benefit of working in the needlet frame -- we can simply and almost optimally combine observations with differing depths and footprints. This feature is only visible in the noise-dominated tSZ map and not the signal-dominated CMB. The continuity of the CMB signal across this boundary provides a simple check of our method.

The E-mode map, like the temperature anisotropies,  is signal-dominated over a wide range of scales. As expected, the characteristic scale of the visible E-mode pattern is significantly smaller than the $\sim\,$degree scale features in the temperature maps. 

The NILC pipeline does not return error estimates on the maps. Simulations are thus a key means of characterizing uncertainties in analyses using these maps. To facilitate this we provide a suite of simulations of these maps. Two types of simulations are provided: first a small suite of non-Gaussian sky simulations -- built with PySM and the \citet{Stein_2020} and \citet{Sehgal_2010} simulations-- and second a set of Gaussian sky simulations. The former is useful for validating analysis pipelines and checking for biases, whilst the later is best suited to characterizing uncertainties. A detailed description of these simulations is given in Appendix \ref{app:simulations}.

\section{Map Properties}\label{sec:map_properties}
During the generation of these maps we investigate a number of their properties that, when combined with the simulation-based pipeline tests discussed in Appendix \ref{app:simulations}, help provide validation of our analysis methods.

\subsection{Properties of the CMB maps}\label{sec:CMB_map_properties}
In Fig.~\ref{fig:cl_tt}, we show the power spectrum of the output temperature map, where for comparison we also show the power spectrum of the equivalent \textit{Planck} \textsc{smica} \citep{planck2016-l04} and our harmonic ACT \& \textit{Planck} ILC maps. The power spectra are computed using \textsc{NaMaster}~\citep{Alonso_2019}. We use bin widths of $\delta \ell=10$ with uniform weight and the error bars are computed analytically with the Gaussian approximation \citep{Wandelt_2001,Efstathiou_2004}. Over a broad range of scales, we see statistical agreement between our ACT and \textit{Planck} map with the \textit{Planck} \textsc{smica} map. On the largest angular scales, we find strong agreement with the \textit{Planck} \textsc{smica} results, as expected as the data are very similar. On the smallest scales we see the dramatic improvement gained by using the small scale ACT data -- at $\ell\sim 2000$ the noise, seen as an upturn in the power spectra, is an order of magnitude lower. Although difficult to determine from the figure, the noise level in the harmonic ILC (HILC) map is approximately 10\% larger on the smallest scales.

\begin{figure}[!ht]
    \centering
  \subfloat[Standard ILC] {\label{fig:stacked_cmb_baseILC} \includegraphics[width=0.48\textwidth]{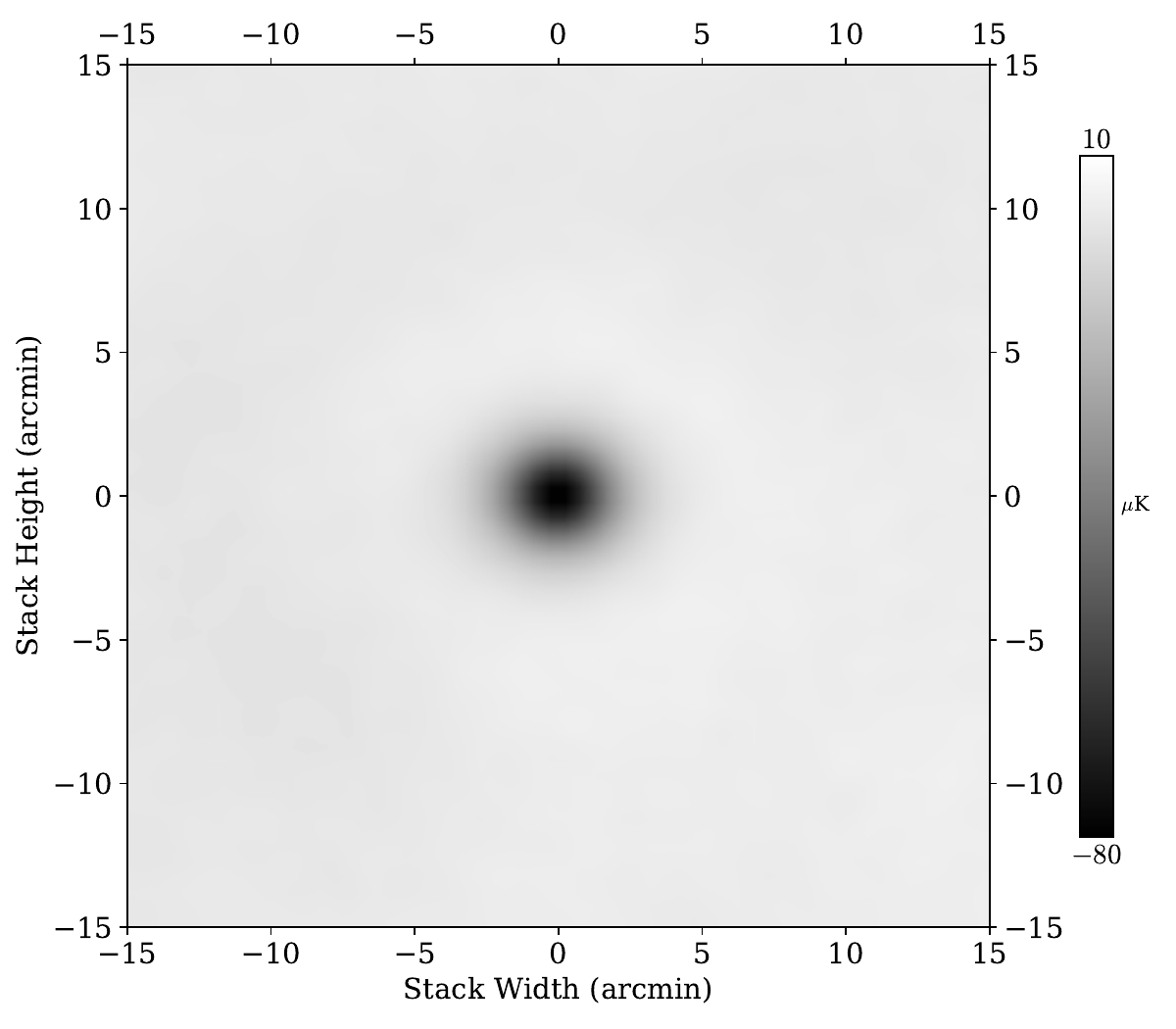} }%
  \\
   \subfloat[ILC with tSZ deprojection] {\label{fig:stacked_cmb_deprojILC} \includegraphics[width=0.48\textwidth]{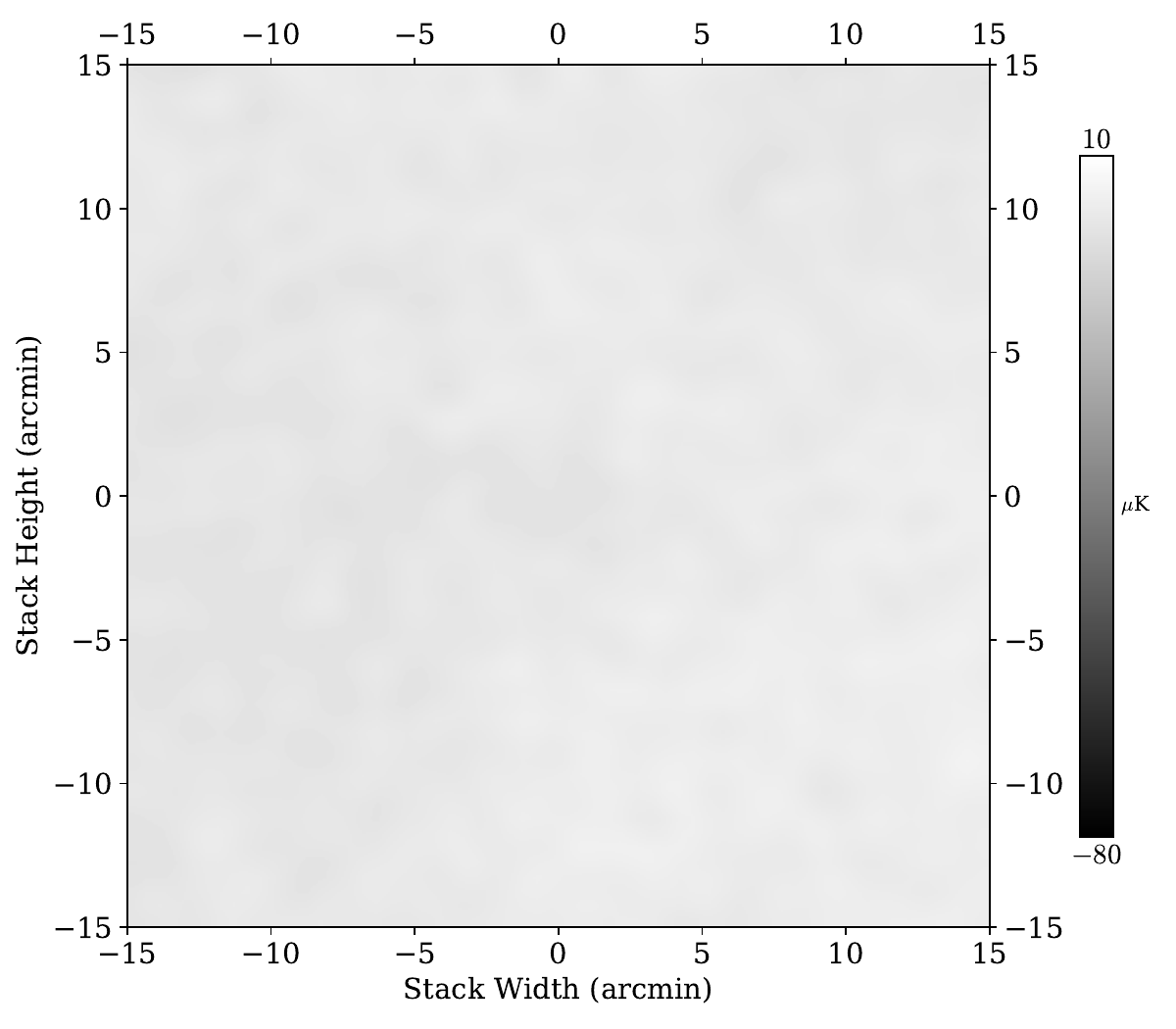}}%
 \caption{ Stacks of the CMB temperature anisotropies at the location of tSZ-detected clusters. The stacks are performed on 30\,arcmin $\times$ 30\,arcmin extracts at the location of clusters in the ACT DR5 cluster catalog \citep{Hilton_2021}. If the ILC perfectly removed all the foregrounds from the CMB map we would expect just noise in these stacks. In the base ILC, Fig.~\ref{fig:stacked_cmb_baseILC}, a large signal caused by residual tSZ contamination is visible. This can be mitigated through the use of a constrained ILC, Section \ref{sec:ILC_intro}, as seen in Fig.~\ref{fig:stacked_cmb_deprojILC}.\label{fig:stacked_cmb_cutouts} }
\end{figure}

As can be seen in Section \ref{sec:maps}, the CMB temperature maps still have residual tSZ contamination. Following Ref.~\citep{Madhavacheril_2018}, these residuals can be seen more explicitly by stacking the CMB map at the locations of clusters detected with ACT DR5 data \citep{Hilton_2021}. Fig.~\ref{fig:stacked_cmb_baseILC} shows the results of this stacking and a large residual of the tSZ signal can be seen. This residual tSZ signal can bias certain analyses \citep[e.g.,][]{Osborne_2014,vanEngelen_2014}. Using the deprojected ILC, as detailed in Section \ref{sec:ILC_intro}, we can create a CMB map with explicitly zero contribution from the tSZ effect. We then perform the same stacking operation and the results are shown in Fig.~\ref{fig:stacked_cmb_deprojILC}. Here we see that the residual is almost completely removed by this procedure. We do see a hint of a small residual signal left at the center of the stack that indicates remaining contamination. A possible source of this contamination is the CIB. CIB galaxies are spatially correlated with the tSZ effect and by stacking on the location of tSZ clusters we are also stacking on CIB galaxies. Further, as is discussed in Ref.~\citep{Sailer_2021,Abylkairov_2021,Kusiak_2023} and Ref.~\citep{Coulton_2022}, when deprojecting the tSZ effect there is often an enhancement of the residual CIB contamination. These two effects combined are thought to give rise to the small residual signal. 

ACT has undertaken a blinding procedure for several key science products, detailed in an upcoming paper, including a blinding of the E-mode power spectrum. This is to minimize biases in the inferred cosmological parameters from effects such as confirmation bias. As such we have not yet performed a power spectrum comparison of the CMB E-mode map to \textit{Planck} data. As is visible in the maps in Fig.~\ref{fig:e_maps}, we have significant improvements in the E-mode noise. A quantified version of the improvement can be seen in Fig.~\ref{fig:cl_EE_sim} of Appendix \ref{app:GaussianSims}, where we show the E-mode power spectra of simulated ILC maps. The data are expected to show similar improvements, and will be assessed in a future paper.

\begin{figure*}
    \centering
  \subfloat[\textit{Planck} MILCA  Compton-$y$ map]{\label{fig:cutout_nilc_planckonly}%
  \includegraphics[width=0.45\textwidth]{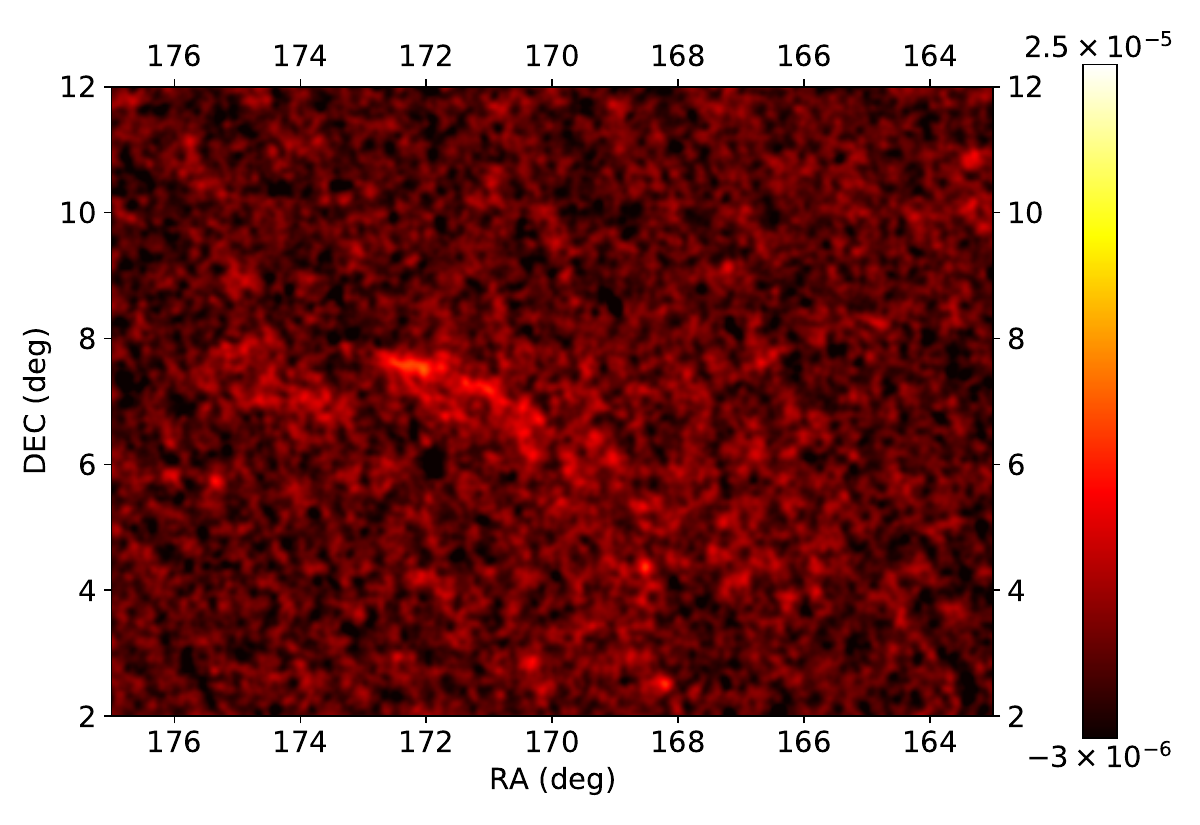}%
 }
  \subfloat[ACT \& \textit{Planck}  NILC  Compton-$y$ map]{\label{fig:cutout_nilc}%
  \includegraphics[width=0.45\textwidth]{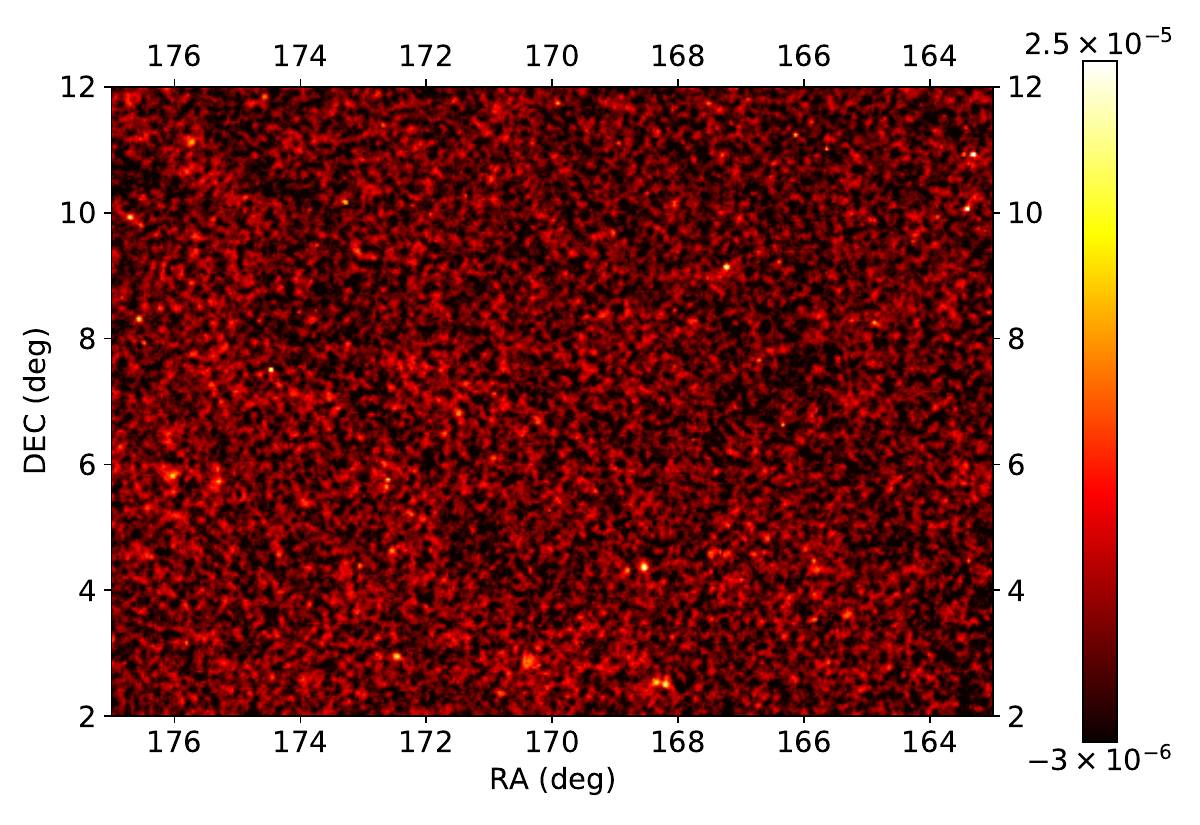}%
 }
 \\ 
  \subfloat[ACT \& \textit{Planck} Harmonic ILC  Compton-$y$ map]{\label{fig:cutout_hilc}%
  \includegraphics[width=0.45\textwidth]{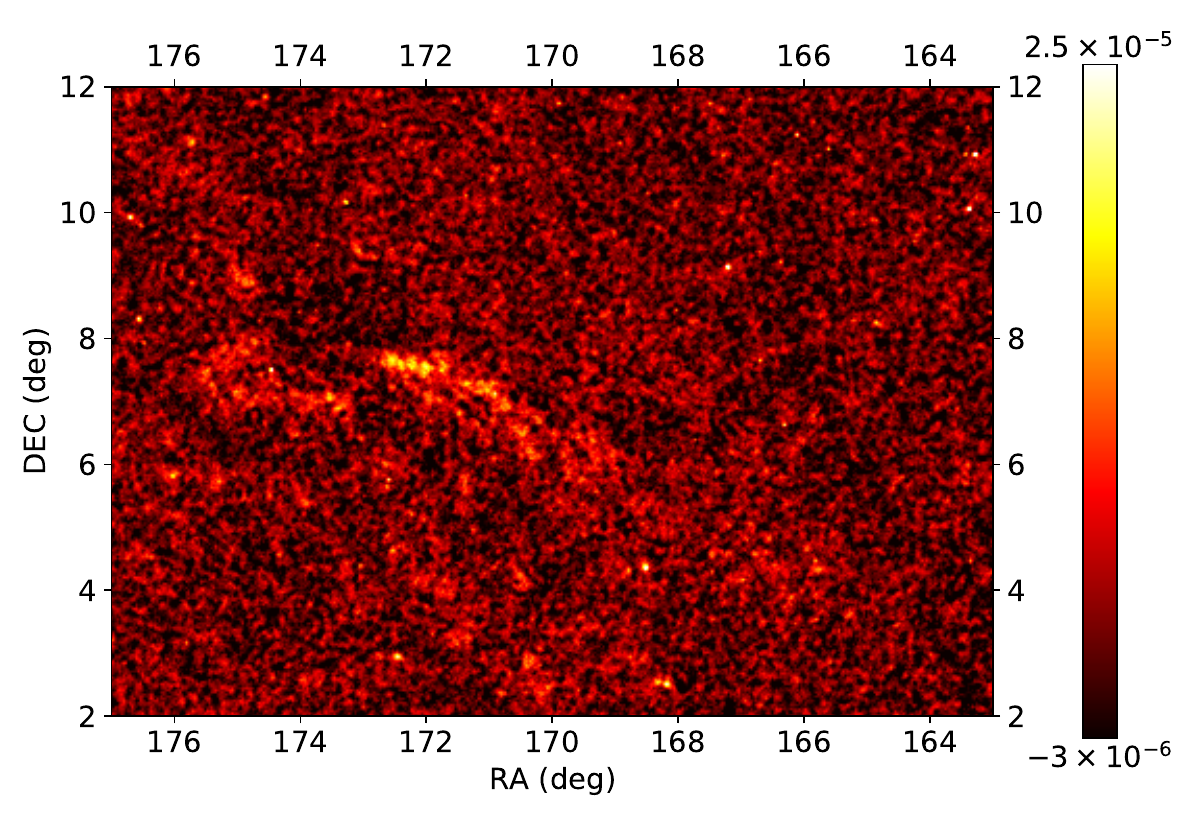}%
 }
  \subfloat[\textit{Planck} $545$ GHz map ]{\label{fig:cutout_545}%
  \includegraphics[width=0.45\textwidth]{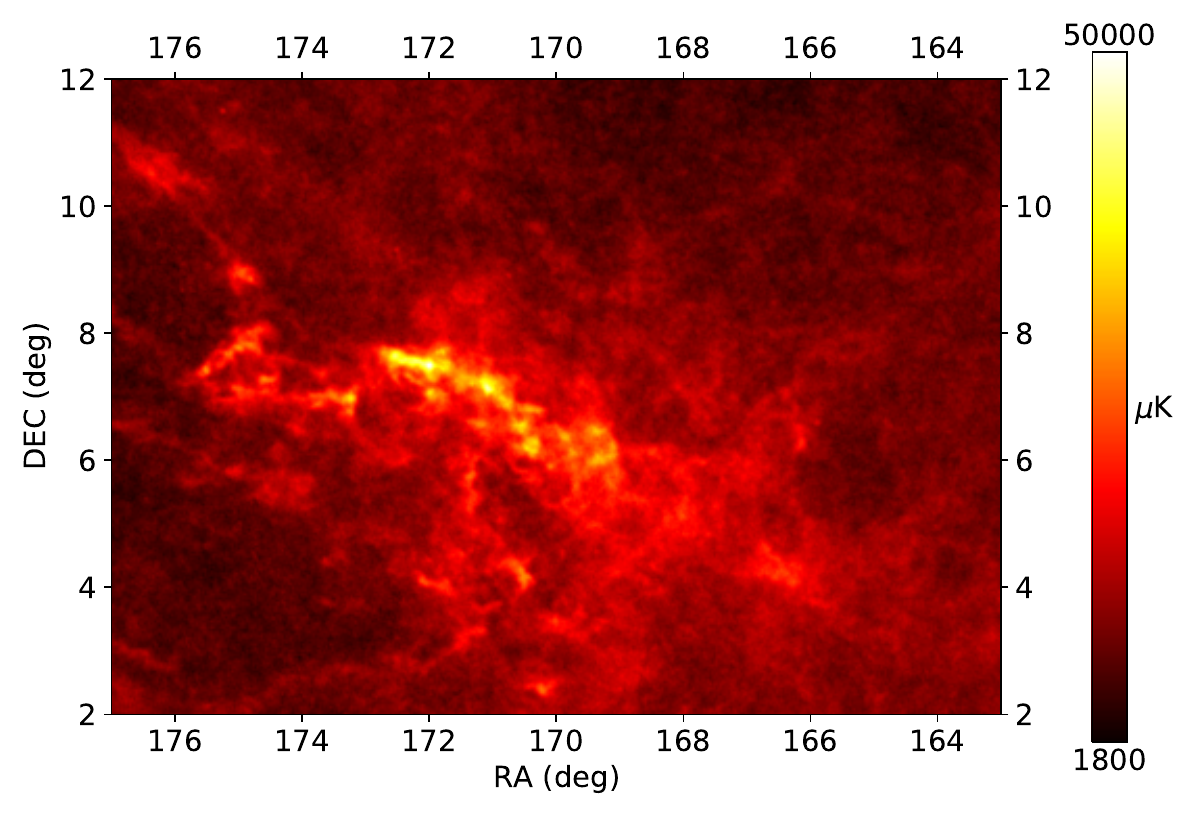}%
 }
 \caption{A comparison of the combined ACT and \textit{Planck} NILC map, the \textit{Planck} 2015 MILCA  map, and a combined \textit{Planck} \& ACT harmonic ILC map of the Compton-$y$ signal on a small $\sim 135$\,deg$^2$ region of sky. To aid the interpretation of these maps we also show the \textit{Planck} 545 GHz map in the same patch of sky. The comparison between the \textit{Planck} MILCA and the ACT \& \textit{Planck} NILC maps demonstrates both the consistency of our results and the benefits of the high-resolution ACT data. The comparison between the harmonic ILC map, the NILC map, and the \textit{Planck} 545 GHz map shows that working in the needlet frame helps remove the strongly spatially varying Galactic foregrounds. Note that to aid visualization the colorbar saturates.\label{fig:SZ_cutout_comp} }
\end{figure*}

\begin{figure}
    \centering
  \includegraphics[width=0.48\textwidth]{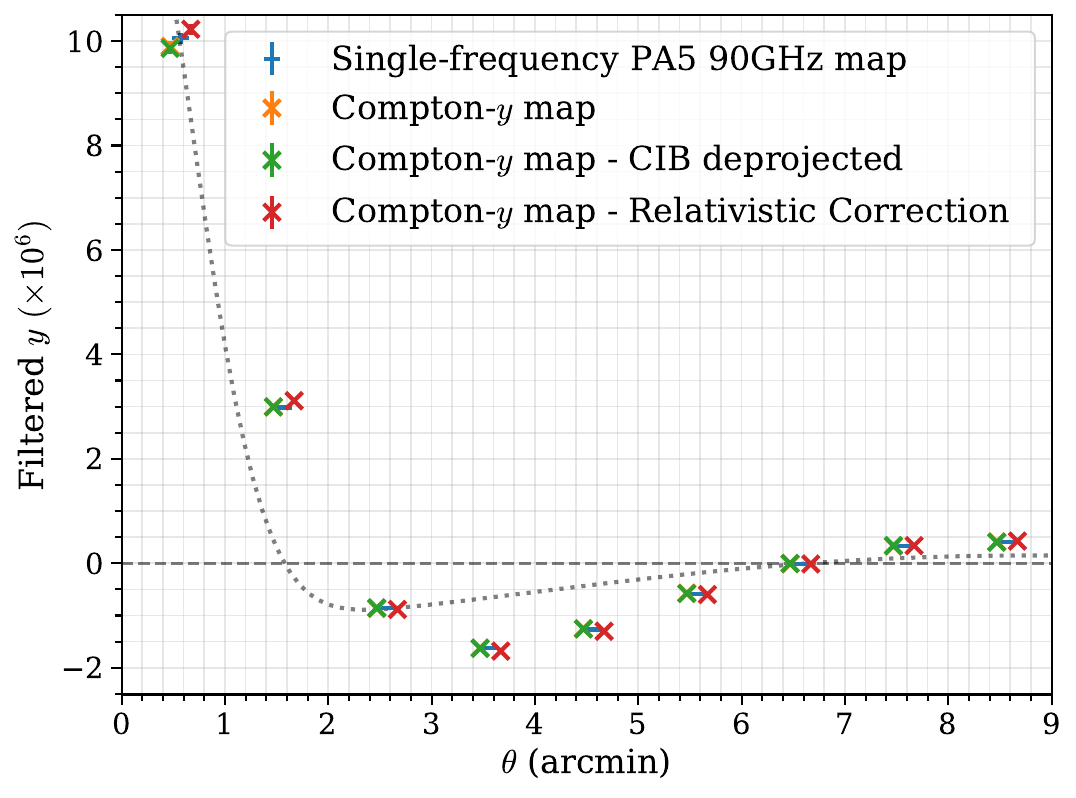}%
 \caption{1D profiles from high-pass filtered stacks of the Compton-$y$ map at the location of tSZ-detected clusters. We perform this stacking operation on four maps: the f$090$ maps from the PA6  detector array, the Compton-$y$ map produced with the non-relativistic tSZ response, the Compton-$y$ map produced with the non-relativistic tSZ response and with the CIB deprojected, and the Compton-$y$ map produced with the relativistic tSZ response. The points with and without the CIB deprojection lie on-top of each other, indicating the measurement is unlikely to be contaminated by CIB emission. Note that the PA5 f$090$ maps has been rescaled into Compton-$y$ units. See Section \ref{sec:freqResponseFunc} for a detailed discussion of the relativistic and non-relativistic responses. We apply a high-pass filter to remove correlated noise, as described in Section \ref{sec:PropComptonY}. To guide the eye, in dashed black we show the scale of the $y$-map beam (1.6 arcmin FWHM) filtered in the same manner as the data. For visualization purposes we have offset the horizontal points.\label{fig:stacked-compton-y} }
\end{figure}

\begin{figure}
    \centering
  \includegraphics[width=0.48\textwidth]{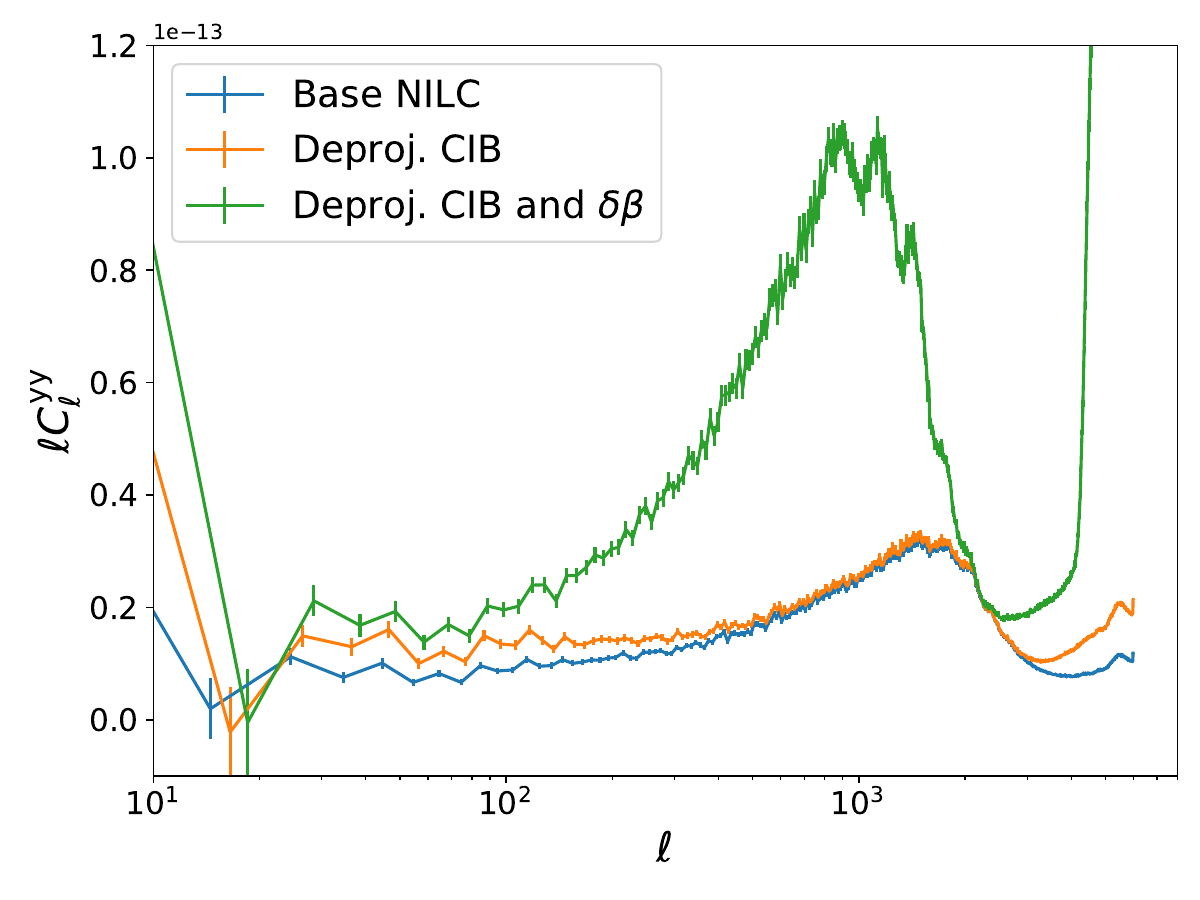}%
 \caption{Power spectra of three different versions of the Compton-$y$ ILC map: the base ILC, a map with a fiducial CIB spectrum explicitly removed, and a map where both a fiducial CIB spectrum and a term to account for uncertainties in the CIB SED, labeled by $\delta \beta$ and described by Eq.~\ref{eq:derivBeta}, are removed. Removing these signals provides increasing levels of robustness to contamination but at a cost of increasing noise, seen here as the increase in power on most scales. The power spectra and error bars are computed in the same manner as those in Fig.~\ref{fig:map_power_spectra}. \label{fig:ILCnoiseCurves} }
 \end{figure}
\subsection{Properties of the Compton-$y$ map}\label{sec:PropComptonY}

The benefits of the high-resolution data and the needlet basis can be most easily seen by examining the Compton-$y$ map as, unlike the CMB map, the Compton-$y$ map is noise-dominated on all scales. This is demonstrated in Fig.~\ref{fig:cl_yy} where we plot the power spectrum of the $y$-map and, for comparison, the expected theoretical thermal Sunyaev-Zel'dovich power spectrum from \textsc{class-sz} \citep{Bolliet_2022}, the \textit{Planck} NILC tSZ map power spectrum \citep{planck2014-a28}, and the harmonic ILC map power spectra. The theoretical model is the same as that used in the Websky simulations \citep{Stein_2020}. All the measurements lie above the theoretical expectation due to the noise bias in these autospectra. The ACT and \textit{Planck} NILC map shows lower noise on all scales compared to the \textit{Planck} NILC map. On large scales this difference arises as the NPIPE maps have lower noise compared to the \textit{Planck}-2015 maps used to compute the \textit{Planck} NILC map. On smaller scales the noticeable improvement comes from the small-scale ACT measurements.\footnote{The conservative Galactic mask used in this work means that the HILC noise is comparable to the NILC noise. If a larger sky fraction were considered the HILC noise would be dramatically larger, as demonstrated in \citep{Chandran_2023}}

Even at the map level many of these benefits are visible, as is demonstrated in Fig.~\ref{fig:SZ_cutout_comp}. First, Fig.~\ref{fig:cutout_nilc_planckonly} and Fig.~\ref{fig:cutout_nilc} demonstrate the benefits of the high-resolution ACT data -- the clusters, visible as bright yellow, point-source-like objects are both better localized and more numerous, demonstrating the depth of the combined $y$-map.  Next, it is beneficial to compare the needlet ILC, Fig.~\ref{fig:cutout_nilc}, to the harmonic ILC, Fig.~\ref{fig:cutout_hilc}. In Section \ref{sec:maps} we saw that the needlets easily accounted for spatial variations in the map depth, and in the harmonic and needlet ILC comparison we can explicitly see how the needlet frame aids the reduction of spatially varying foregrounds. In the center of the harmonic ILC patch, a bright extended structure is visible, but  it is suppressed in the needlet ILC maps. This structure is the imprint of residual Galactic dust, as can be seen in Fig.~\ref{fig:cutout_545} where we show this region as observed by the dust dominated \textit{Planck} 545 GHz channel. Through the localization in real-space the needlets are able to treat low dust and high dust regions differently. On the other hand, the harmonic ILC can only operate on the sky-averaged properties. 

Examining the profiles of the Compton-$y$ clusters is a powerful test of the Compton-$y$ map. In Fig.~\ref{fig:stacked-compton-y} we compare the average profile of tSZ clusters as measured in the Compton-$y$ map and in one of the input f$090$ maps, stacked using the ACT DR5 cluster catalog \citep{Hilton_2021}. We apply a high-pass filter to remove modes with $\ell<2000$, which are `noise' dominated and contribute correlated noise to the stack. We expect the profiles of the clusters at f$090$ to be similar to the clusters in the Compton-$y$ map, after accounting for the different units and beams. Thus the good agreement seen here demonstrates that our pipeline is not distorting the cluster profiles and amplitudes. Note that, as ACT and \textit{Planck} are not sensitive to the monopole, the monopole of the Compton-$y$ map is not physically meaningful. Thus the Compton-$y$ map is a map of the fluctuations of the Compton-$y$ field about the mean.

The high sensitivity and large area of these maps means that we need to take particular care to account for potential biases. A key bias in Compton-$y$ maps is the cosmic infrared background (CIB). The dusty star-forming galaxies, which source the CIB, are spatially correlated with the tSZ effect as some of these galaxies occupy the massive halos that source the tSZ effect. Hints of CIB contamination have been seen in previous tSZ analyses \citep[see e.g.][]{planck2014-a29,Schaan_2021,Vavagiakis_2021}. As discussed in Section \ref{sec:component_separation} we can minimize the impact of the CIB by explicitly removing any signal with a specified modified blackbody spectrum and, for cases that are especially sensitive to CIB biases, implementing an additional correction for deviations around this modified blackbody, as described in Section \ref{sec:freqResponseFunc}. The cost of this additional removal is increased noise that can be seen in Fig.~\ref{fig:ILCnoiseCurves}, where we compare the power spectra of the base Compton-$y$ map to maps with these different deprojections.

Generally, each additional component deprojected results in a further increase of the noise in the ILC map. This is seen as an increase in power in the power spectrum as the maps are noise-dominated. There are several noteworthy features in the deprojected maps: firstly the noise penalty of the deprojections is highest on scales where only ACT data contributes (the smallest scales). This arises because ACT lacks the very high \textit{Planck} frequency channels that have strong sensitivity to the CIB. Without these channels it is difficult to separate the tSZ and CIB, hence the large noise increase. The noise starts to increase around $\ell\sim 3000$ as this is where the high frequency \textit{Planck} data becomes noise-dominated. This increase is more significant when more components are deprojected. Interestingly on the scales with $1400 \lesssim \ell \lesssim 2500$, there is no noise penalty for deprojecting just the CIB and, in a smaller range of scales, no cost for deprojecting both the CIB and the correction term.  This occurs on scales where the noise in the ILC map is dominated by non-CIB contributions. In this regime, noise in the CIB deprojected ILC maps is not set by the ability to separate the CIB and tSZ components, but instead the noise arising from the other components (in this case atmospheric noise and residual CMB).

\begin{figure}
    \centering
  \includegraphics[width=0.48\textwidth]{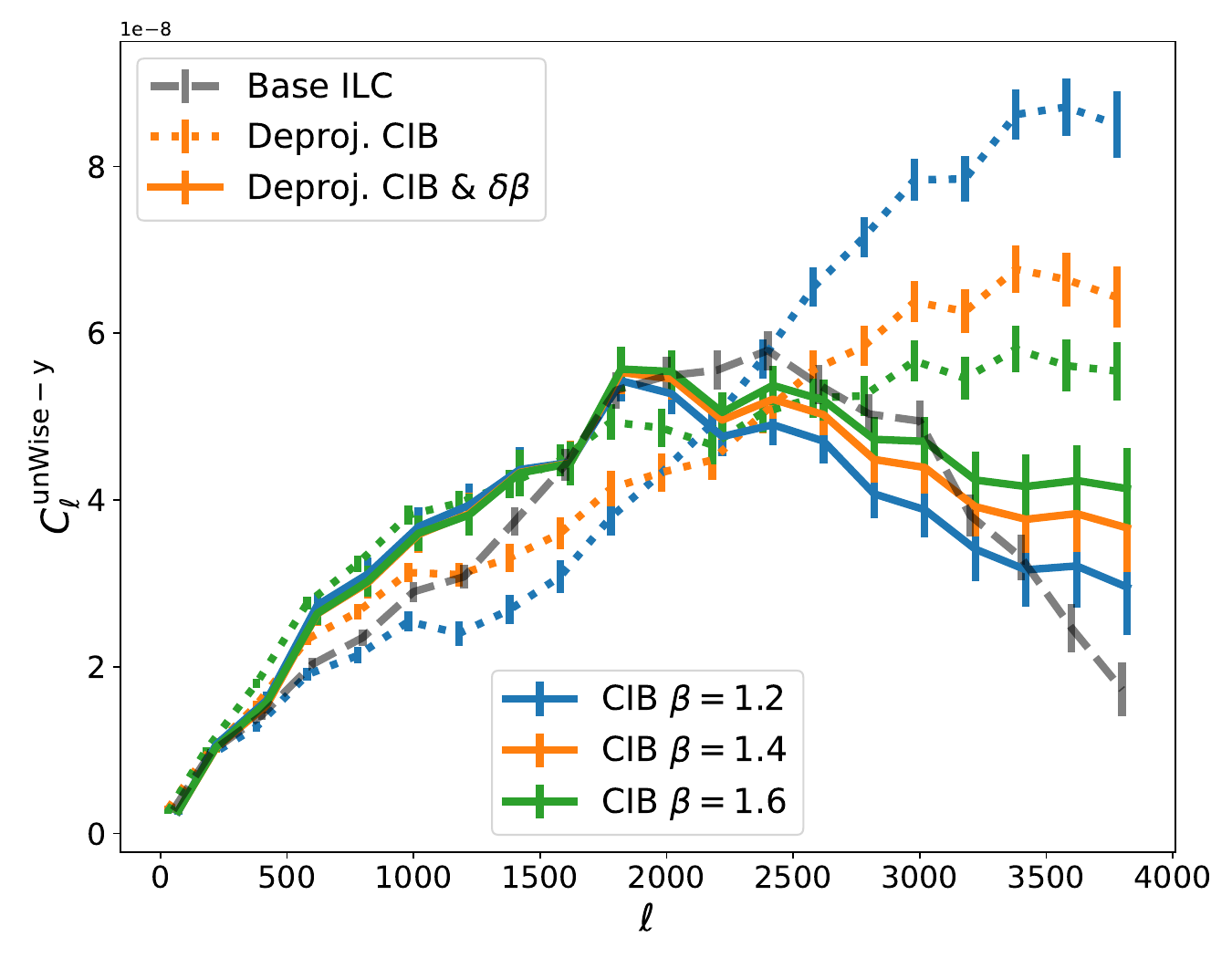}%
 \caption{ A preview of preliminary results from Ref.~\citep{Kusiak_2023b} -- an upcoming analysis of correlations between the tSZ effect and the \emph{unWISE} galaxies. Here we show the cross-correlation of the Compton-$y$ map with the \emph{unWISE} “blue” subsample, of mean redshift $z=0.6$. The measured Compton-$y$ -- \emph{unWISE} power spectrum shows evidence of CIB contamination, as seen in the difference between the measurement on ILC maps with and without CIB deprojection. Due to the strong physical correlation between the \emph{unWISE} and CIB galaxies, this analysis is also sensitive to how the CIB is removed and choosing different parameters for the CIB spectral index leads to different results. One method of mitigating this is to also deproject a correction to the CIB SED, given in Eq.~\ref{eq:derivBeta}. This term helps correct for any mismatch between the true SED and the assumed model. As shown by the solid lines, this approach leads to results that are more robust to modelling choices. Note both the \emph{unWISE} and Compton-$y$ maps are dimensionless. \label{fig:cl_unwise_y} }
 \end{figure}

As an example of how these maps can be used in an analysis, we highlight the workflow used in an upcoming cross-correlation analysis of ACT and \emph{unWISE} galaxies \citep{Kusiak_2023b}. A component of this analysis is the cross-power spectrum between the ACT \& \textit{Planck} Compton-$y$ map and the \emph{unWISE} galaxy catalog \citep{Meisner_2017, Krolewski_2020, Krolewski_2021, Kusiak_HOD_2022}. The \emph{unWISE} galaxies are highly correlated with the CIB \citep{Kusiak_2023}. To test the sensitivity of this analysis to the CIB, we compare power spectra measurements from the base NILC $y$ map with the CIB-deprojected $y$ map. As can be seen in Fig.~\ref{fig:cl_unwise_y}, where we show the results from \citep{Kusiak_2023b} for the \emph{unWISE} “blue” subsample (of mean redshift $z=0.6$), deprojecting the CIB leads to a significant shift in the measurement. Further, when different spectral indices $\beta$ are assumed for the deprojected CIB, statistically significant shifts are observed. This suggests that this analysis is strongly sensitive to contamination from CIB such that the CIB SED needs to be carefully modeled. One means of mitigating this is to use the moment expansion method, discussed in Section \ref{sec:freqResponseFunc}, and deproject the derivative term as well. As can be seen in Fig.~\ref{fig:cl_unwise_y}, the use of the moment expansion leads to results that are more robust to the choice of CIB parameters. The small differences between the true CIB SED and the assumed model are absorbed by the correction term, and all curves with the moment expansion deprojection, the solid lines in Fig.~\ref{fig:cl_unwise_y}, converge around the same values for $\ell\lesssim 2500$. The efficacy of the moment expansion method in accounting for uncertainties in the spectral properties of a contaminant has also been seen in Ref.~\citep{Azzoni_2021}, where they use the same approach to remove Galactic dust emission. As a second example, consider the stacked profiles shown in Fig. \ref{fig:stacked-compton-y}. We can minimize any potential bias by performing the stack on a CIB-deprojected map. However, the points are essentially unchanged, as seen in Fig. \ref{fig:stacked-compton-y}, and this result is stable to variations in the CIB spectral index. This suggests that for that analysis CIB contamination is less severe and there is no need to deproject the CIB or the derivative. The difference in sensitivity arises as these two example analyses are sensitive to halos of very different mass and redshift. 
 
Finally we compare the results of using the non-relativistic and relativistic tSZ responses. Using the non-relativistic response can bias the resulting $y$-map as is discussed in Ref.~\citep{Hurier_2017,Erler_2018} and Ref.~\citep{Remazeilles_2020}. The importance of this difference can be most directly seen in the tSZ cluster stacks -- in Fig.~\ref{fig:stacked-compton-y} we compare stacks of detected galaxy clusters in a Compton-$y$ map made using the non-relativistic temperature response, Eq.~\ref{eq:tSZResponse}, and the average temperature response, Eq.~\ref{eq:resp_tsz_scale_dep}. We see that the profiles in the latter case are $\sim 5\%$ larger, with the difference arising as the standard tSZ map is biased low by the ignored relativistic SZ contributions.

\section{Conclusions}\label{sec:discussion} 
\begin{table*}
\begin{tabular}{c |  c | c}
Sky Component & Deprojected components & Notes  \\ \hline
\multirow{3}{15em}{\shortstack[c]{CMB temperature \\ and kinetic Sunyaev-Zel'dovich} } & tSZ &  \\
&CIB (T$_\mathrm{CIB}=10.7$\,K, $\beta=1.7$) & $\ell_\mathrm{max}=17000$\\
&tSZ \& CIB &\\
\hline
CMB E-mode  & None &  $\ell_\mathrm{max}=4000$ \\
\hline
\multirow{3}{15em}{\shortstack[c]{Compton-$y$ \\ relativistic and non-relativistic} } &CIB &\\
& CIB \& CIB correction  & $\ell_\mathrm{max}=17000$ \footnote{The $\ell_\mathrm{max}$ of maps with two deprojections is reduced to $\ell_\mathrm{max}=11000$}  \\
& CIB \& CMB & \\
\end{tabular}
\caption{A summary of the maps to be delivered. In addition to the minimum-variance ILC map, we produce CMB temperature and Compton-$y$ maps that deproject one or more components. All maps are convolved to a 1.6 arcmin FWHM Gaussian beam. \label{table:products} }
\centering
\end{table*}

We have presented component-separated CMB temperature, CMB E-mode polarization, and Compton-$y$ maps that trace the integrated gas pressure.  These maps were produced with a needlet ILC pipeline, designed to take advantage of the different localization of key map properties. To mitigate the well-known ``ILC bias" we developed a simple scheme that helps ensure that the resulting ILC maps have no significant loss of the signal.

In addition to the ILC bias, we explored biases arising from residual foreground contamination in the ``cleaned" ILC maps. These residuals are left from the imperfect separation of sky signals. The importance of these residuals depends on the analysis in question, but for many analyses these residuals can cause important biases in cosmological and astrophysical inferences.  For such analyses we have created a set of constrained ILC maps, as described in Ref.~\citep{Remazeilles_2013}, that have one or more foregrounds explicitly removed -- at the cost of extra noise. The derived products with different deprojections are summarized in Table \ref{table:products}. It is important to note that the Fourier space filtering used in this work removes fewer modes than previous ACT analyses. This means that there may be small scan-synchronous pickup residuals in the maps. For cross-correlation studies this can be safely ignored, but other analyses should perform tests to ensure their results are not impacted by residual scan-synchronous pickup. Further the Fourier space correction introduces a small amount of anisotropic noise.For most analyses this can be ignored, but if it is important then optimal filtering routines \citep[e.g.][]{Wandelt_2004,Smith_2007} that incorporate the anisotropic structure of the noise can be used to maximally utilize these maps. 

Finally, there are additional observational systematic effects that can impact this analysis. As discussed in Ref.~\citep{Madhavacheril_2020} and Ref.~\citep{Jason_2010}, atmospheric transmission, calibration errors and passband uncertainties can alter the signal seen at each frequency. The passband and gain uncertainties for ACT result in {$\sim 1.5\%$ variations in the amplitude of the tSZ signal in each channel (note that f220 has larger variations, which is partly driven by the null of the tSZ effect at $217\,$GHz and partly by the larger uncertainty)}. In Appendix \ref{app:systematics} we explore the impact of these systematic effects in detail and find that the uncertainties in the passbands, beams, and calibrations lead to negligible changes to stacked cluster profiles from the Compton-$y$ map. Whilst this result cannot easily be related to the precise impact of these instrumental effects on other scientific analyses, it suggests the size of these effects in the Compton-$y$ map are small.

The NILC maps isolate the component of interest whilst minimizing noise and, particularly when using deprojections, contaminants. This makes these ideal for a broad set of science cases; previous component-separated ACT maps have been used for analyses ranging from detailed studies of individual clusters and filaments \citep{Hincks_2022}, to studies of galaxy group and cluster astrophysics with large ensembles \citep{Amodeo_2021,Schaan_2021,Kusiak_2021,Gatti_2022,Lokken_2022}, to studies of the distribution of matter with lensing \citep{Darwish_2021}. The maps presented here will enable the statistical power of such analyses to be significantly increased -- the noise in the Compton-$y$ map presented here is similar to that of the deeper ``D56" map from Ref.~\citep{Madhavacheril_2020} but over an area of sky that is $\sim 25\times$ larger. Likewise, the large improvement in resolution over the \textit{Planck} component separated maps, seen in Fig.~\ref{fig:map_power_spectra}, would be highly beneficial to the many analyses based on these maps \citep[e.g.][]{Hill_2014,Hurier_2015,planck2014-a29,planck2017-LIII,planck2016-l09,Carron_2022}. 

It is important to note that there are classes of analysis for which the individual frequency maps would be more appropriate to use; this includes those that require high-precision characterization of the NILC noise. For example,  analyses of the power spectrum of the primary CMB anisotropies are likely best done with the frequency maps as modelling the foregrounds with a parametric model enables explicit marginalization over the foregrounds. Furthermore, that approach facilitates fine-grained modeling of instrumental systematics and noise. Precisely modeling the power spectrum of the NILC noise and propagating the associated uncertainties, which would be necessary to robustly extract the primary CMB contribution from the maps presented here, would be equivalent to modelling the frequency channels separately, with the added complication of propagating these components through the NILC pipeline. Thus, the analysis of the power spectrum of the primary CMB anisotropies for ACT DR6 is expected to be done with the frequency maps.  Similarly these maps were not used in the recent ACT DR6 lensing analysis. Whilst component-separated maps have been used in past lensing analyses \citep{planck2016-l08,Carron_2022}, the challenges in dealing with the complex ACT noise (discussed in Ref. \citep{Atkins_2022,Qu_2022}) motivated a simpler analysis of the individual frequency maps. 

Finally, these maps overlap with numerous ongoing surveys, such as the Dark Energy Survey \citep{DES_2016}, Hyper Suprime-Cam \citep{HSC_2018} survey, and Dark Energy Spectroscopic Instrument \citep{DESI_2016} survey, and therefore are very well-suited for a range of cross-correlation studies. The pipeline developed in this work is highly versatile and can be used to map other sky signals or be applied to other data sets, including upcoming CMB missions such as the Simons Observatory \citep{SO_2019} or CMB-S4 \citep{S4_2016}.

\begin{acknowledgments}
The authors are grateful to Reijo Keskitalo for help with the NPIPE products, Jens Chluba, Mathieu Remazeilles, and Aditya Rotti  for useful discussions.  Computing was performed using the Princeton Research Computing resources at Princeton University, the National Energy Research Scientific Computing Center (NERSC),  the Niagara supercomputer at the SciNet HPC Consortium and the Rusty cluster at the Flatiron Institue. SciNet is funded by the CFI under the auspices of Compute
Canada, the Government of Ontario, the Ontario Research Fund–Research Excellence, and the University of Toronto. 

Support for ACT was through the U.S.~National Science Foundation through awards AST-0408698, AST-0965625, and AST-1440226 for the ACT project, as well as awards PHY-0355328, PHY-0855887 and PHY-1214379. Funding was also provided by Princeton University, the University of Pennsylvania, and a Canada Foundation for Innovation (CFI) award to UBC. ACT operated in the Parque Astron\'omico Atacama in northern Chile under the auspices of the Agencia Nacional de Investigaci\'on y Desarrollo (ANID). The development of multichroic detectors and lenses was supported by NASA grants NNX13AE56G and NNX14AB58G. Detector research at NIST was supported by the NIST Innovations in Measurement Science program. 

MM, AL acknowledge support from NASA grant 21-ATP21-0145. BDS, FJQ, BB, IAC, GSF, NM, DH acknowledge support from the European Research Council (ERC) under the European Union’s Horizon 2020 research and innovation programme (Grant agreement No. 851274). BDS further acknowledges support from an STFC Ernest Rutherford Fellowship. EC, BB, IH, HTJ acknowledge support from the European Research Council (ERC) under the European Union’s Horizon 2020 research and innovation programme (Grant agreement No. 849169). JCH acknowledges support from NSF grant AST-2108536, NASA grants 21-ATP21-0129 and 22-ADAP22-0145, DOE grant DE-SC00233966, the Sloan Foundation, and the Simons Foundation. CS acknowledges support from the Agencia Nacional de Investigaci\'on y Desarrollo (ANID) through FONDECYT grant no.\ 11191125 and BASAL project FB210003. RD acknowledges support from ANID BASAL project FB210003. ADH acknowledges support from the Sutton Family Chair in Science, Christianity and Cultures and from the Faculty of Arts and Science, University of Toronto. JD, ZA and ES acknowledge support from NSF grant AST-2108126. KM acknowledges support from the National Research Foundation of South Africa. AM and NS acknowledge support from NSF award number AST-1907657. IAC acknowledges support from Fundaci\'on Mauricio y Carlota Botton. LP acknowledges support from the Misrahi and Wilkinson funds. MHi acknowledges support from the National Research Foundation of South Africa (grant no. 137975). SN acknowledges support from a grant from the Simons Foundation (CCA 918271, PBL). CHC acknowledges FONDECYT Postdoc fellowship 3220255. AC acknowledges support from the STFC (grant numbers ST/N000927/1, ST/S000623/1 and ST/X006387/1). RD acknowledges support from the NSF Graduate Research Fellowship Program under Grant No.\ DGE-2039656.
OD acknowledges support from SNSF Eccellenza Professorial Fellowship (No. 186879). CS acknowledges support from the Agencia Nacional de Investigaci\'on y Desarrollo (ANID) through FONDECYT grant no.\ 11191125 and BASAL project FB210003. TN acknowledges support from JSPS KAKENHI Grant No.\ JP20H05859 and No.\ JP22K03682. AvE acknowledges support from NASA grants 22-ADAP22-0149 and 22-ADAP22-0150. JC acknowledges support from a SNSF Eccellenza Professorial Fellowship (No. 186879).

This work made use of the \textsc{numpy}\citep{numpy}, \textsc{scipy}\citep{scipy}, \textsc{healpy}\citep{healpy}, \textsc{astropy}\citep{astropy:2013}, \textsc{h5py}\citep{h5py}, \textsc{pyfftw}\citep{pyfftw}, \textsc{libsharp}\citep{libsharp} and \textsc{matplotlib}\citep{matplotlib} libraries.
\end{acknowledgments}
\newpage
\appendix

\section{Simulations}\label{app:simulations}
Simulations are a key part of many analyses and so we also provide simulated component-separated maps. We provide two main simulation products: simulated non-Gaussian sky maps and Gaussian simulations, using the methods developed in Ref.~\citep{Atkins_2022}. These two products allow tests of different aspects of analysis pipelines: the non-Gaussian simulations are ideal for studying the properties of fields like the Compton-$y$ maps, where signal non-Gaussianity is highly important, whilst the Gaussian simulations are useful for computing ensemble quantities that require large numbers of simulations.

Specifically, the available products are: 
\begin{itemize}
    \item Four different non-Gaussian realizations of the ACT region of the sky (two from the Websky and two from the Sehgal et al. simulation suites). Each sky realization is provided with instrumental and atmospheric noise simulated via two complementary models - a wavelet model and a tiled model.
    \item Wavelet and tiled noise models, with associated code, that can be used to generate Gaussian simulations of the `ILC' noise in the blackbody and Compton-$y$ maps. 
    \item  A small number of end-to-end Gaussian sims of the blackbody and Compton-$y$ maps -- 10 each for the tiled and wavelet models. These are the products of processing Gaussian simulations of the input data sets through the NILC pipeline.
\end{itemize}
All of these products are produced with $\ell_\mathrm{max}=10000$.

\subsection{Non-Gaussian Simulations}\label{app:EndToEndnonGaussianSims}
These simulations are composed of three components: Galactic signals, extragalactic signals, and instrumental and atmospheric noise. We generate a non-Gaussian realization of the Galactic sky signals using \textsc{PySM}; we include free-free, synchrotron, thermal dust and Anomalous Microwave Emission (AME) in temperature and the thermal dust and synchrotron signals in polarization \citep{Thorne_2017,Zonca_2021}.\footnote{ We use models ``d1", ``s1", ``a1" and ``f1".}

For the extragalactic signals we use two different suites of simulations: the Websky simulations and the Sehgal et al simulations \citep{Li_2022,Stein_2020,Sehgal_2010}. These two suites both provide one full-sky realization of the lensed CMB sky, the thermal and kinetic Sunyaev-Zel'dovich effects, the CIB, and radio galaxies. The two suites of simulations assume different physical models for each of the components. We refer the reader to Ref.~\citep{Sehgal_2010} and Ref.~\citep{Stein_2020} for the detailed differences of the modelling. When compared, these simulations provide some measure of the importance of our modeling uncertainties. Unfortunately each simulation only has a single realization, which can limit some studies. However, as the ACT footprint covers only $\sim 1/3$ of the sky, we can rotate the sky and construct a second simulation of the ACT footprint from these simulations. Specifically, we generate the first realization by cutting out the ACT footprint from the full-sky simulation and generate the second realization by first rotating the simulations by 90$^\circ$ and then cutting out the ACT footprint. Note that this cannot easily be done for the Galaxy due to its spatial anisotropy.

The final component is simulations of the noise. For \textit{Planck} we use the end-to-end NPIPE noise simulations \citep{planck2014-a14}. For ACT we generate simulations using the models described in Ref.~\citep{Atkins_2022}; one set using the tiled noise model and a second set using the wavelet noise model. By providing simulations from both of the noise models, it will be possible to test how the uncertainties in our noise modelling impact each analysis.

Bringing these pieces together, for each map in our data set we generate a simulated sky by combining the PySM Galaxy with either the Websky or Sehgal et al. simulations at the same frequency. These maps are then convolved with the appropriate instrument beam. Next we add instrumental noise to this to complete the mock observation. We repeat this operation for each map and then input these maps into the needlet ILC pipeline. Note that in these simulations we use Dirac delta function passbands, i.e., the simulations are evaluated at each single frequency, and the ACT maps here have no simulated transfer function. 

\begin{figure}
    \centering
  \includegraphics[width=0.48\textwidth]{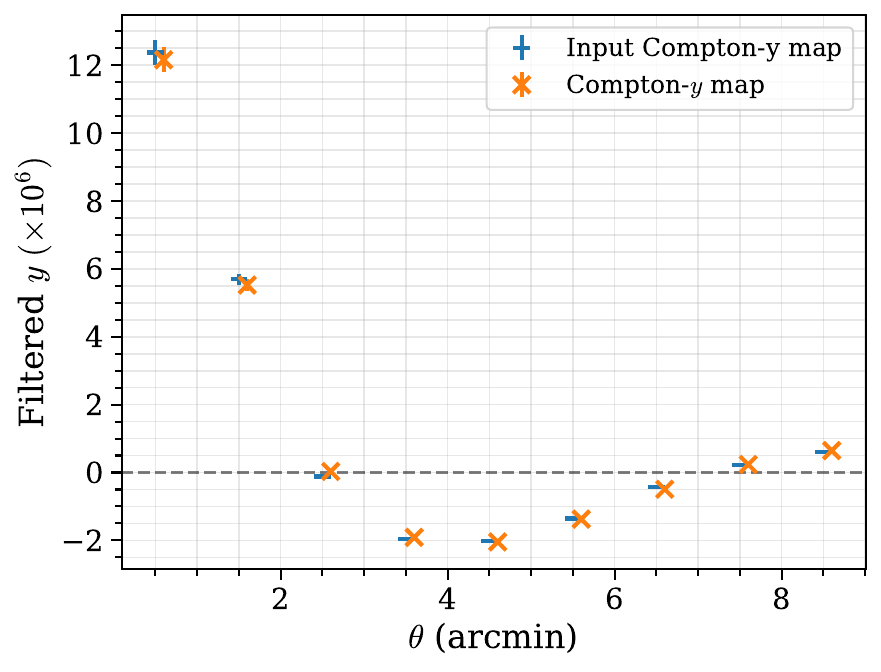}%
 \caption{A validation of the Compton-$y$ map with the Websky simulations. We compare the stacked NILC Compton-$y$ map with the true input Compton-$y$ map. The stack is performed on all halos with $M>5 \times 10^{14}$ M$_\odot$/$h$ and these maps are high-pass filtered in the same manner as the data in Fig.~\ref{fig:stacked-compton-y}. \label{fig:stackWebskySim} }
\end{figure}

As a first use case of these simulations, we compare the stacked profile of tSZ clusters in the input map to those in the output needlet ILC Compton-$y$ map, with the results shown in Fig.~\ref{fig:stackWebskySim}. We find very good agreement between the two sets of stacked profiles and thereby validate this aspect of our pipeline. A benefit of having simulations is that we can analyze each component of the map separately, exploiting the linearity of the ILC method. Thus, we can explore how just the CIB component contributes to the final tSZ map. These maps are useful for understanding residual signals in the maps and for assessing potential biases.

These simulations have a few key limitations: firstly, we only have four sky realizations, given by the two rotations of the two simulations.  Secondly, the sky power of these simulations does not match the data -- whilst the Websky and Sehgal et al simulations are sophisticated models of the sky, they do not perfectly capture all the components, leading to differences in the size of each sky component. For example, the small scale power arising from the discrete nature of the CIB and radio galaxies is not accurate. Additionally, there are minor limitations (e.g., delta function passbands and simple beam treatments) that prevent tests of certain systematic effects. Despite these limitations, they are invaluable for analyses of non-Gaussian aspects of the sky.
\subsection{Gaussian Simulations}\label{app:GaussianSims}
To complement the non-Gaussian simulations we provide noise models as described in Ref.~\citep{Atkins_2022}. These allow the fast generation of many realizations of the noise in the needlet ILC maps and are ideal  for investigating ensemble properties of the ILC map's noise.

We generate these noise models using a two stage process: first, we obtain a model of the sky components present in the input maps. For this purpose we use the Ref.~\citep{Dunkley_2013} model, extended to include power laws for the Galactic dust, synchrotron, Anomalous Microwave Emission, and free-free, and a white noise component of arbitrary amplitude and correlation across the maps to account for the Poisson contribution of point sources. The latter component is necessary as we remove all detected sources in each map, rather than those down to a specific flux cut, and this results in each map having a slightly different level of Poisson power. The strength of this correlation across the input maps varies due to the different ratio of radio to dusty sources in each map. Further, we modulate the Galactic dust components by a smoothed version of the \textit{Planck} dust intensity map. Without this modulation the large scale power in the component separated maps is significantly below that of the data, highlighting how the  spatial variations of the Galaxy complicate component separation.

Second, we use this sky model to simulate observations of the sky. We use the same methods as in Appendix \ref{app:EndToEndnonGaussianSims} to simulate the instrumental noise and beams. We then process these simulations with the needlet ILC pipeline. From the resulting needlet ILC maps, we subtract the input realization of the signal of interest, e.g., for the CMB ILC maps we subtract the CMB power; this gives a map of the NILC `noise'. Using the tools developed in Ref.~\citep{Atkins_2022} we then construct tiled and wavelet noise models. These noise models allow arbitary numbers of realizations of the NILC `noise' to be generated. The noise realizations can then be combined with realizations of the signal to provide a simulated map.

\begin{figure}
    \centering
  \includegraphics[width=0.48\textwidth]{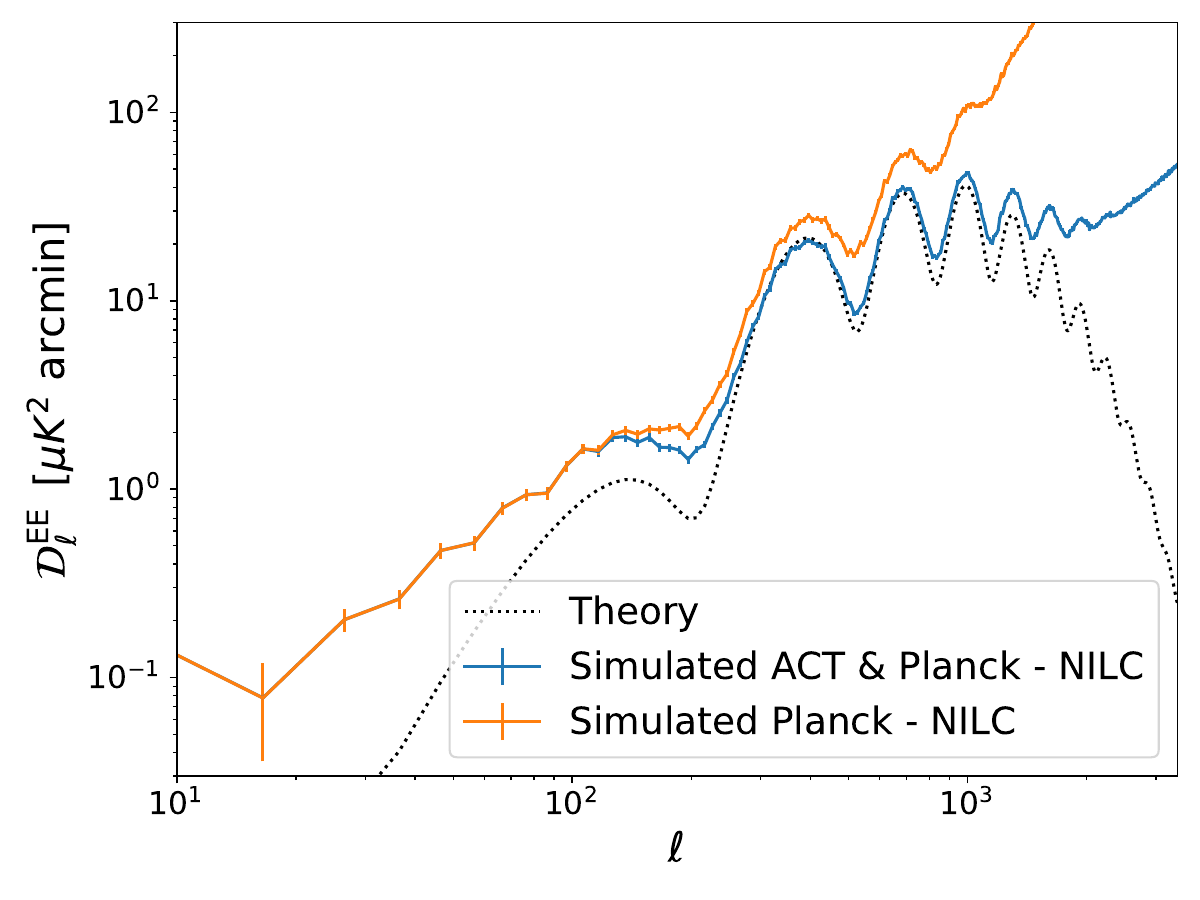}%
 \caption{The E-mode auto power spectrum of a simulated \textit{Planck}-only ILC map and a simulated \textit{Planck} \& ACT ILC map. It can clearly be seen that the ACT data allow the E-modes to be probed to significantly smaller scales. \label{fig:cl_EE_sim}}
\end{figure}

The end-to-end Gaussian simulations produced in generating the noise model allow further tests of potential biases in our pipeline.  As discussed in Section \ref{sec:filteringCorrection} and seen in Fig.~\ref{fig:cls_biased_vs_debiased}, these simulations can be used to assess the level of bias in our ILC maps. We find no significant biases across the full range of scales. In Fig.~\ref{fig:cl_EE_sim} we compare the power spectra of simulated \textit{Planck}-only to \textit{Planck} \& ACT ILC E-mode maps. This demonstrates the value of combining the ACT data with \textit{Planck}.

\section{Harmonic ILC}\label{app:HarmonicILC}
Instead of working in the needlet domain, the ILC method can be applied in the harmonic domain -- hereafter the harmonic ILC. In this work we use the harmonic ILC as a baseline to compare against the NILC results, and to build intuition for the bias reduction methods, as the harmonic frame is conceptually simpler. In this Appendix we outline the harmonic ILC method and describe the harmonic-domain implementation of our ILC bias mitigation method.

For the harmonic ILC we have
\begin{align}
\hat{s}_{\ell m} = \sum\limits_i w_i a^{i}_{\ell m}
\end{align}
where the weights are defined in Eq.~\ref{eq:basic_ilc_weights}. The empirical covariance matrix, $\hat{\mathcal{C}}$, required to compute the weights is simply the estimated power spectra
\begin{align}\label{eq:covMatHILC}
\hat{\mathcal{C}}_{ij} =\frac{1}{N_\mathrm{modes}} \sum\limits_{L-\delta\ell<\ell \leq L+\delta\ell}\sum\limits_m a^{i}_{\ell m}{a^{j}}^*_{\ell m},
\end{align}
where we compute the power spectrum over a range of $\ell$, quantified by 2$\delta\ell$ and $N_\mathrm{modes}$ is the number of modes in each bin. We average over a small range of $\ell$ to improve the estimate of the covariance matrix and minimize the ILC bias. To further minimize ILC bias we introduce an ILC bias reduction method, in a similar manner to those discussed in Section \ref{sec:ILCBias}. When computing the ILC weights for mode $m$ we exclude that $m$ mode from the covariance matrix estimation, Eq.~\ref{eq:covMatHILC}. To avoid the computational cost of estimating the covariance matrix $\ell$ times per scale, we instead split the modes into $N=10$ subsets (eg., the first $\ell /N$ $m$ modes are assigned to subset 1 etc) and, for each subset, estimate the covariance matrix with all modes not in that subset. This allows the ILC bias to be mitigated with reduced computational overhead, at the cost of a slight increase in the resulting ILC map's noise. In fact, whilst the standard method of estimating the covariance matrix has lower noise, it is biased low. When we correct for this bias, by dividing out a transfer function, we tend to find almost identical noise levels to those of our method. 

When computing harmonic ILC maps we use the same steps as the needlet ILC pipeline, described in Section \ref{sec:pipeline}, including the frequency dependent responses. 

\begin{figure}
    \centering
  \includegraphics[width=0.48\textwidth]{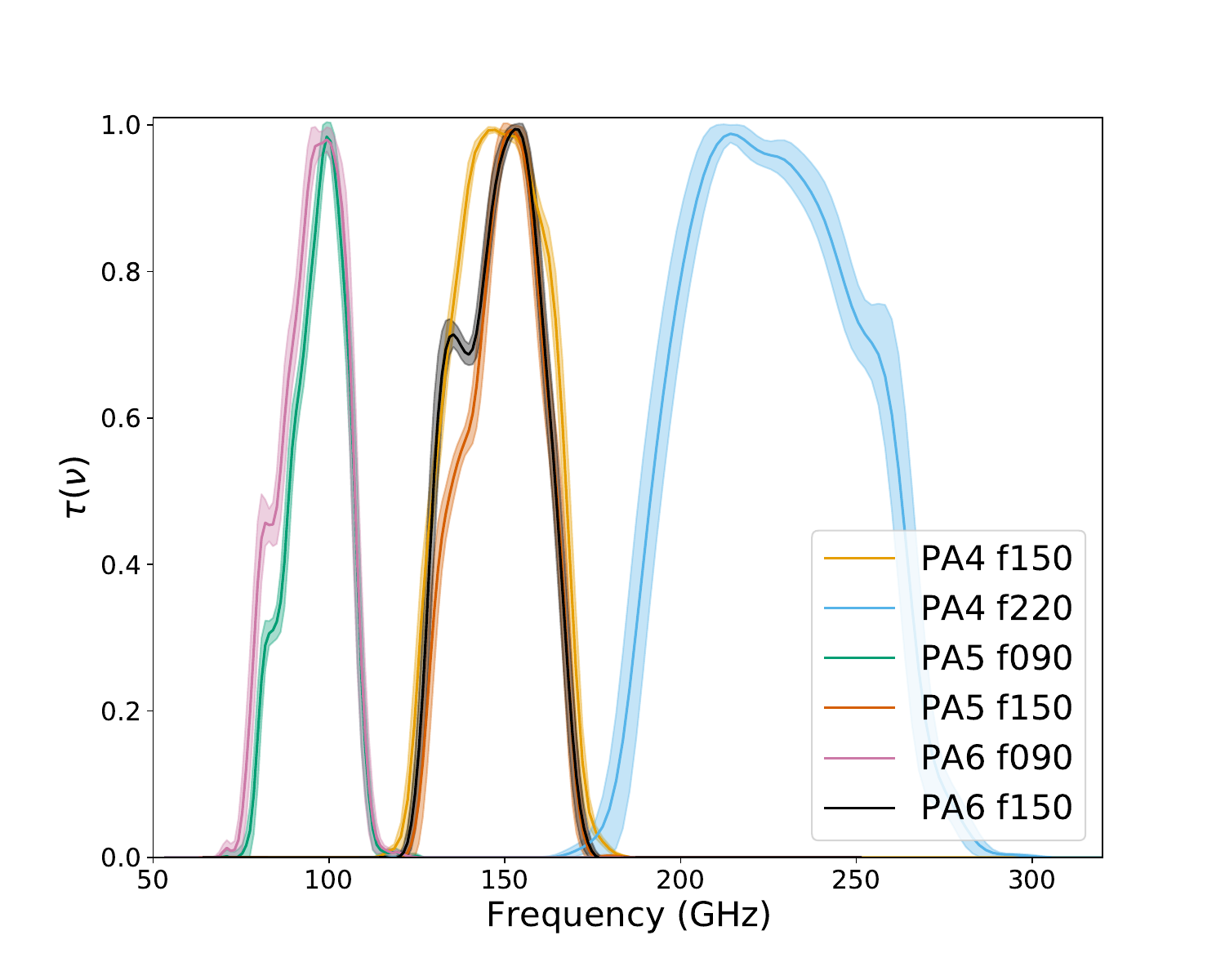}%
 \caption{The passbands for the six different ACT DR6 data sets. These are used in  Eq.~\ref{eq:responseFunc} to compute the responses to each sky signal. The shaded region denotes the $1\,\sigma$ uncertainty on these measurements. \label{fig:passbands_wErrors}}
\end{figure}

\section{Mitigating the large scale ILC bias}\label{app:needlet_m_modes}

As described in Section \ref{sec:ILCBias}, we mitigate the large scale ILC bias by isolating a subset of the modes from the needlet band and using the remainder to compute the ILC weights. In detail this isolation can be performed as follows: starting from the map of anisotropies at a given needlet scale, we perform a SHT. We apply a filter that selects a subset of $a_{\ell m}$ modes and then we perform a second SHT to transform these modes back to the needlet frame; we call this the `data map'. This `data map' is then subtracted from the original needlet map.\footnote{For computational efficiency we combined these two steps with the initial transformation into the needlet frame. This saves one SHT transform. } This new map contains all the modes except the held out modes and we use this map to compute the `covariance maps' and ILC weights. The ILC weights are then applied to the `data maps'. This procedure is repeated with the next subset of $a_{\ell m}$ modes etc. The resulting ILC maps from the subsets are then added back together to obtain the ILC solution for that needlet scale.

Whilst the approach of removing a subset of $a_{\ell m}$ modes is very similar to the harmonic ILC, described in Appendix \ref{app:HarmonicILC}, there is a slight additional subtlety in the needlet frame. When we remove $a_{\ell m}$ modes from the needlet band, the localization of the needlets is altered. If we were to systematically remove each single $a_{\ell m}$ mode the change in the localization of the covariance matrix would be negligible. In practice this would not be effective as effects such as masking and non-Gaussianites mean that each $a_{\ell m}$ mode is not independent. Thus this would only reduce, but not eliminate, the ILC bias. Further, removing modes from the needlet basis requires spherical harmonic transforms, and so this operation comes at a significantly increased computational cost. Hence, this motivates our choice to remove small subsets of $a_{\ell m}$ modes across the band.

The decision of how many subsets to use, $N_\mathrm{subset}$, and how to divide the modes into these subsets was made by ensuring the localization of the covariance matrix is not strongly altered. This is balanced by computational time (more subsets is slower) and stability (changing masks and power spectra in simulations should not dramatically change the ILC maps). We use $N_\mathrm{subset}=5$. For our footprint the strong correlations from the mask led us to assign $\ell<4$ to subset 1, $4<\ell<8$ to subset 2 etc up to $\ell=20$. Beyond $\ell=20$, the assignment is based on the $m$-mode at each $\ell$. Modes with $0\leq m<\ell/N_\mathrm{subset}$ are assigned to subset $\ell/20$ MOD $N_\mathrm{subset}$. Modes with $\ell/N_\mathrm{subset}\leq m< 2 \ell/N_\mathrm{subset}$ are assigned to subset $1+\ell/20$ MOD $N_\mathrm{subset}$ etc. The adjacent $m$-modes are assigned to the same subset as these are most tightly coupled by effects such as masking. Further, by removing different $m$-modes for each $\ell$ we ensure that the localization of the new covariance matrix closely approximates the original covariance matrix. This is additionally aided by our smoothing operation and the fact that we remove only a small subset of the total modes in the needlet band.  One might worry that the small set of modes that the covariance matrix is applied to correspond to some highly non-local feature in the maps. However, this is not an issue due to the linearity of the ILC method: in the standard ILC identical results are obtained if the ILC weights are applied mode-by-mode or applied to all modes simultaneously. Instead it is important that the covariance matrix, and hence the weights, remains highly localized and we ensure this with the choices described above.

\section{Instrumental Systematic Effects}\label{app:systematics}

\begin{figure}
    \centering
  \includegraphics[width=0.48\textwidth]{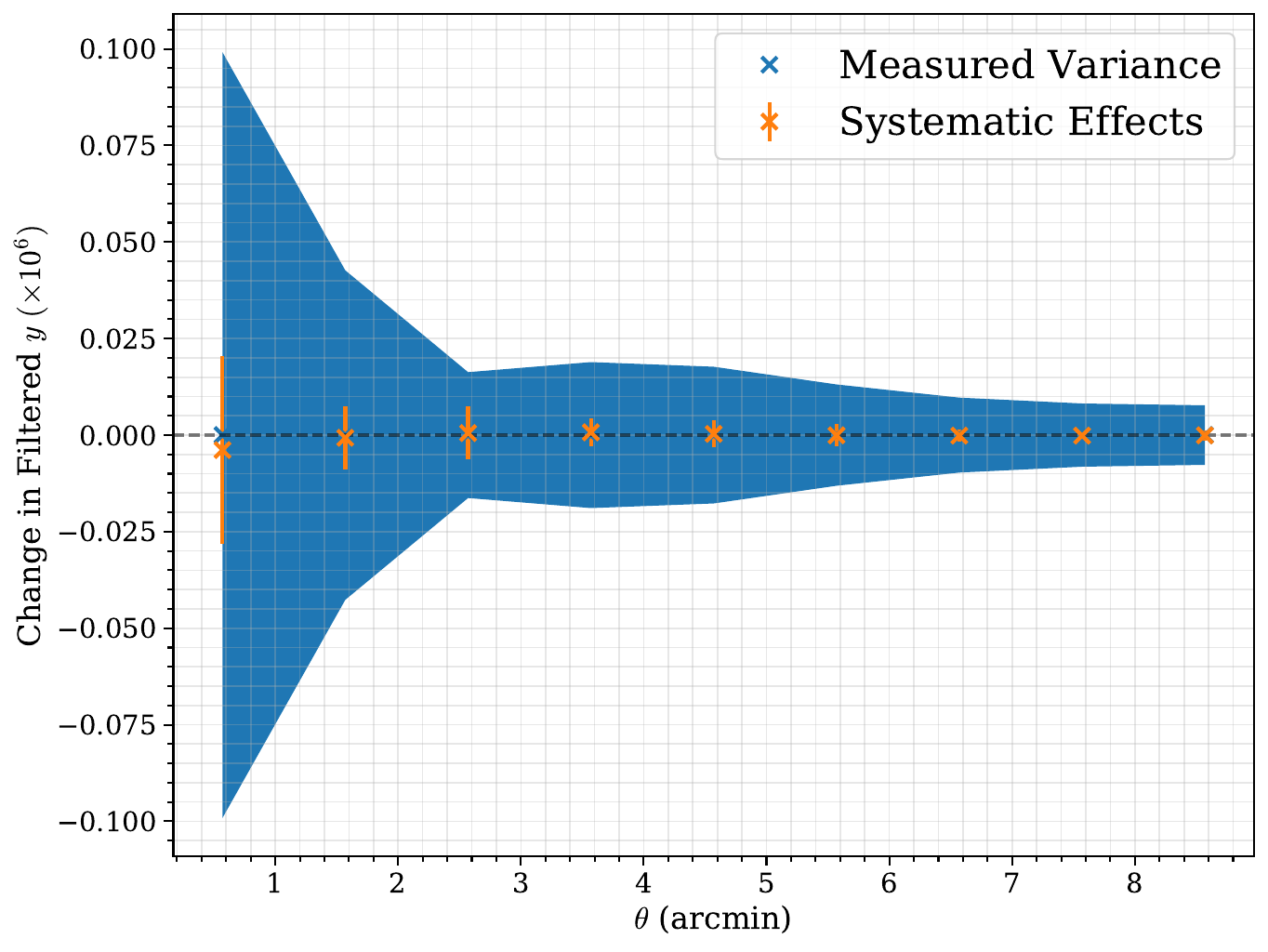}%
 \caption{The absolute change in the stacked cluster profiles, Fig.~\ref{fig:stacked-compton-y}, caused by the calibration, beam, and passband uncertainties. These effects, shown in orange, are significantly smaller than the measured scatter in the stack, shown as a blue band. \label{fig:systematics}}
\end{figure}

The key instrumental systematic effects that impact our analyses arise from the miscalibration of the input maps and uncertainties in the passbands and beams. All of these effects can be expressed as changes to the ILC responses, ${r}_\nu$ in Section \ref{sec:ILC_intro}, and are important as they control how maps are combined to extract the signal of interest. If these are misestimated the resulting ILC map can be biased. In this section we investigate how significant these biases can be.

The impact of a miscalibrated map is simply a rescaling of the ILC responses $\mathbf{r}_\nu \rightarrow (1+\delta c) \mathbf{r}_\nu$, where $\delta c$ is the amplitude of the miscalibration. Note that the miscalibration factor is the same for every signal on the sky. On the other hand, the passband misestimation affects each sky component differently. The impact on each of the responses can be assessed by recomputing them with altered passbands, i.e., changing $\tau(\nu)$ in Eq.~\ref{eq:scaleDepResp}. Beam uncertainty has two impacts: firstly, it acts as a scale dependent miscalibration, given by the ratio of the estimated beam to the true beam. Secondly, it alters the color correction to the response, given in Eq.~\ref{eq:scaleDepResp}.

To assess the importance of these terms we sample calibration factors, passbands, and beams within their uncertainties. For the \textit{Planck} passbands we shift the central frequencies based on the uncertainties reported in Ref.~\citep{planck2020-LVII} and use the calibration errors reported in Ref.~\citep{planck2020-LVII}. The \textit{Planck} beam errors have been shown to be negligible \citep{planck2016-l05} and so are neglected here.  For ACT DR4 we use the same calibration and beam uncertainties as are included in the power spectrum analyses \citep{Louis_2017,Choi_2020,Lungu_2022} and use the passband errors from Ref.~\citep{Madhavacheril_2020}. For DR6 errors we use similar procedures to compute the equivalent errors, with the details of this provided in an upcoming ACT paper. The passbands used in this work, with their errors, are shown in Fig.~\ref{fig:passbands_wErrors}.

In Fig.~\ref{fig:systematics} we show the impact these effects have on the stacked Compton-$y$ profiles. No systematic shift is observed and the spread caused by these effects is negligible. 

This result does not ensure that instrumental systematic effects are negligible for other science use cases. In particular the beam uncertainties are scale dependent (and are generally larger on small scales) and the passband uncertainties will likely become more important in analyses that require multiple deprojections.  Instead, if instrumental effects are a concern, this analysis should be repeated. Despite these limitations, these results can be used to estimate the size of the effect.

\bibliographystyle{apsrev.bst}
\bibliography{nilc,planck_bib}

\end{document}